\documentclass[12pt]{elsarticle}

\usepackage{comment}
\usepackage{hyperref}
\usepackage{times}
\usepackage{physics}
\usepackage{graphicx}
\usepackage{miller}
\usepackage{color}

\newcommand{\cplusa}{$\langle\textrm{{\bf c}}+\textrm{{\bf a}}\rangle$}

\newcommand{\deltaEpyr}{$\Delta E^{\textrm{I-II}}_{\textrm{Mg}}$}

\newcommand{\sigmaFluc}{$\sigma\left(\mathcal{F}\left[\Delta E^{\textrm{I-II}}_{\textrm{Mg-x}}(l_{CXS})\right]\right)$}

\newcommand{\deltaEpyrmacro}{$\Delta E^{\textrm{I-II}}_{\textrm{Mg}}(\boldsymbol{\epsilon})$}

\newcommand{\Us}{$U_j$}
\newcommand{\deltaUs}{$\Delta U_j^{\textrm{I-II}}$}

\newcommand{\deltaUpyr}{$\Delta U_j^{\textrm{I-II}}$}

\newcommand{\DFTFE}{\texttt{DFT-FE}~}

\begin{document} 
\begin{frontmatter}



\title{Intrinsic ductility enhancement in Mg alloys elucidated via large-scale ab-initio calculations}



\author[a]{Sambit Das\fnref{fn1}}
\author[a,b]{Vikram Gavini\corref{author}}
\cortext[author] {Corresponding author.\\\indent \textit{E-mail address:} vikramg@umich.edu}
\address[a]{Department of Mechanical Engineering, University of Michigan, Ann Arbor, MI 48109, USA}
\address[b]{Department of Materials Science \& Engineering, University of Michigan, Ann Arbor, MI 48109, USA}

\begin{abstract}
Magnesium is the lightest structural alloy, yet its practical use is limited by its low ductility. Recent studies suggest  ductility enhancement in dilute Mg alloys may stem from favorable solute modification of \cplusa~pyramidal I/II screw dislocation core energy difference, activating \cplusa~slip via a double cross-slip mechanism. This work conducts large-scale DFT calculations, reaching $\sim$6,000 atoms, of \cplusa~dislocation energetics in Mg and Mg-Y/Zn alloys. We find that relative solute strengthening effects on pyramidal I and II screw dislocation glide are crucial for cross-slip enhancement in Mg-Y, in contrast to prior investigations, that find solute-mediated dislocation-core energy modification as the main driver. Our predictions align with single- and poly-crystal experimental results and also capture the transition from pyramidal II to I preferred slip in Mg-Y.

\end{abstract}
 
\begin{keyword}
Electronic structure, Pyramidal dislocation, magnesium, ductility, cross-slip, alloy.
\end{keyword}

\end{frontmatter}












\section{Introduction}
Magnesium (Mg) is the lightest structural metal and abundantly available in the earth's crust. The high strength to weight ratio of Mg makes it an ideal candidate for light-weight metallic alloys in automotive and aerospace sectors~\cite{Pollock986}. However, its technological use has been severely limited by the lack of ductility at room temperature. The low ductility of Mg is a consequence of the well known plastic anisotropy in its hexagonal close packed (HCP) crystal structure, accommodating only two independent easy glide slip systems---$\hkl{0 0 0 1}\hkl<1 1 -2 0>$ basal slip (BS)---with a low critical resolved shear stress (CRSS) ($<$1MPa~\cite{tonda2002effect}).  The $\hkl{1 0 -1 1}\hkl<-2 1 1 3>$ first-order (I) pyramidal \cplusa~slip (FPCS) and $\hkl{1 1 -2 2}\hkl<-2 1 1 3>$ second-order (II) pyramidal \cplusa~slip (SPCS) are also glissile slip systems, although with moderate CRSS values (40--60 MPa~\cite{Rikihisa2017}). However, the \cplusa~pyramidal (Pyr) edge dislocation segments are metastable and undergo a thermally assisted transformation to a much lower energy ($\sim$3 eV/nm lower) basal-dissociated sessile dislocation structure at room-temperature~\cite{wu2015origins}.  
Two pathways for improving room temperature ductility of Mg, by accommodating non-basal plastic deformation, constitute deformation twinning~\cite{BARNETT20071,YUE20233427} and \cplusa~slip~\cite{YOO200187}. In this work, we focus on the \cplusa~slip mechanism due to its potential for enabling large plastic strains~\cite{LiuScience2019}, and has been a subject of intense recent investigations~\cite{fan2015towards,cepeda2023activation,WuCurtinScience2018,sandlobes2017rare,Sandlobes2011,HU20222717,NANDY20211521,lentz2016strength}. Specifically, experimental studies have demonstrated that dilute solute alloying of Mg with transition metal solutes  significantly improve ductility of Mg~\cite{Sandlobes2011,SANDLOBES201492}. For instance, even a small concentration ($\sim$0.8 at.\%) of Y solute enhances the strain to fracture by around 3 to 5 times compared to pure Mg~\cite{Sandlobes2011}. Recent optical microscopy based slip trace analysis in Mg-Y single and poly-crystals~\cite{Rikihisa2017,Rikihisa2020} has  associated the enhanced ductility with increased activation of FPCS and SPCS, with FPCS  being more preferred beyond $\sim$0.5 at.\% Y. Mg-Zn is another important binary system with  moderate  ductility enhancement (2--3 times) observed for concentrations under 0.5 at.\% Zn enabled by SPCS~\cite{ando2012plasticMgZn,shi2013effects}. A mechanistic understanding of \cplusa~slip activation in Mg alloys is crucial for computational design of binary and ternary Mg alloys, including rare-earth (RE) free Mg alloys~\cite{sandlobes2017rare}.

The prevailing  understanding of the mechanism driving the enhanced ductility of Mg-RE alloys was put forth in Refs.~\cite{WuCurtinScience2018,AhmadCurtin2019,AhmadCurtin2020}, ascribed to the accelerated double cross-slip of \cplusa~Pyr screw dislocations through dilute solute alloying. The cross-slip events are also observed experimentally in tensile testing of 0.8 at.\% Mg-Y alloy~\cite{WuCurtinScience2018}. In these works, a dislocation line tension based model was used to estimate the cross-slip rate, a measure of intrinsic ductility. The studies suggest that the cross-slip energy barrier ($\Delta G_{XS}$), estimated using the line tension model, reduces with increasing Y concentration (still in the dilute limit), thus enhancing the cross-slip rate and non-basal plastic activity. The line tension model adopted in the study requires two critical inputs from ab-initio electronic structure calculations: (i) the core energy difference between the \cplusa~screw dislocations dissociated on pyramidal I and II planes in pure magnesium (\deltaEpyr$=E^{\textrm {I}}_{\textrm{Mg}}-E^{\textrm{II}}_{\textrm{Mg}}$), and (ii) the interaction energy of these dislocations with substitutional solute atoms, denoted by \Us, with $j$ denoting the site index of a dislocation core site. Accurate calculations of these quantities from density functional theory (DFT) have been challenging due to the large simulation domains needed to compute these quantities that have been out of reach using conventional DFT codes. Notably, the only estimates of \deltaEpyr~are using a quadrapole arrangement of dislocations with 600 atoms~\cite{Itakura2016}, which resulted in significant cell-size errors, and another recent study~\cite{LIU2024119864} that used a larger simulation domain of 2000 atoms but was limited to a Mg pseudopotential with two valence electrons. On the contrary, as will be demonstrated in this study, a large simulation domain in conjunction with an accurate pseudopotential including the semi-core electrons is necessary to accurately compute \deltaEpyr, a quantity with high sensitivity to cross-slip prediction. Further, direct DFT calculations to compute the dislocation solute interaction energies (\Us) require even larger simulation domains to avoid spurious solute-solute interactions from periodic images, and have not been possible thus far. In order to bypass the direct ab-initio computation of these crucial inputs into the line tension model, previous efforts~\cite{WuCurtinScience2018,AhmadCurtin2019} used a surrogate approach involving DFT computed interaction energies of solutes with the Pyr I/II stacking fault and elastic interaction energy between the solute misfit strain tensor and dislocation elastic stress field. The more sensitive \deltaEpyr~quantity was estimated by fitting the ductility enhancement to experimental data~\cite{WuCurtinScience2018}.

In this work, we conduct direct DFT calculations of \deltaEpyr~and \Us~using \DFTFE---a massively parallel real-space DFT code based on adaptive finite-element discretization~\cite{Motamarri2019,das2022dft} enabling fast and accurate large-scale DFT calulations. Notably, we find simulation domains with $\sim$3,000 atoms per periodic layer are needed to obtain a cell-size converged value for \deltaEpyr~to $\sim$ 1 meV/nm accuracy. Moreover, we find domain sizes of $\sim$6,000 atoms are required for cell-size converged \Us~calculations, conducted for the Y and Zn solutes. Importantly, we find that direct DFT calculations of \deltaEpyr~and \Us~do not support the findings in Ref.~\cite{WuCurtinScience2018} where the enhanced cross-slip in Mg-Y alloy was attributed to the relative stabilization of Pyr I with respect to Pyr II \cplusa~screw dislocation with increasing Y concentration. On the contrary, we find \deltaEpyr~is positive and increases with Y concentration, which suggests that Pyr II screw dislocation is increasingly stabilized with respect to Pyr I with increasing Y concentration. Our study reveals the importance of accounting for (i) influence of external macroscopic stress on \deltaEpyr~
, and (ii) solute concentration dependent CRSS of Pyr I and II screw dislocations. Accounting for these effects, using DFT inputs, explains many aspects of experimentally observed enhanced ductility in Mg-Y and Mg-Zn alloys, including the changing preference of dominant \cplusa~slip activity (SPCS/FPCS) with concentration.

\section{Computational method}\label{sec:computationalMethods}
Below we discuss the computational methodology and set up for the large-scale DFT calculations for obtaining the \deltaEpyr~and \Us~inputs to the cross-slip model. The theoretical aspects of their incorporation into the cross-slip model are discussed in Sec.~\ref{sec:incorporation} and the Appendix.
\subsection{General details of the large-scale DFT calculations}\label{sec:largeScaleDFT}
The DFT simulations in this work are performed using the \DFTFE software~\cite{das2022dft,Motamarri2019,Motamarri2018,MOTAMARRI2013308}. \DFTFE  is based on systematically convergent high-order spectral finite-element basis, and implements the Chebyshev filtering acceleration technique~\cite{ChFSISAAD2006,MOTAMARRI2013308} for solution of the Kohn-Sham non-linear eiegenvalue problem. The computational complexity for DFT-FE  scales quadratically with the number of electrons up to system sizes of
30,000 electrons~\cite{Motamarri2019}. \DFTFE is GPU accelerated, enabling fast and accurate pseudopotential DFT calculations on  large-scale systems~\cite{GB2019,GB2023}.
We used the  Perdew-Burke-Ernzerhof  generalized gradient approximation (PBE-GGA) exchange-correlation functional~\cite{pbe}. Further, we employed optimized norm-conserving pseudopotentials~\cite{oncv2013} for Mg, Y and Zn from the Pseudo-Dojo database~\cite{van2018pseudodojo}, and a Fermi-Dirac smearing temperature of 500 K. 

\subsection{Computational methodology for \deltaEpyr~ calculations}\label{sec:deltaEpyrSetup}
\begin{figure}[htpb]
 \centering
 \includegraphics[width=0.8\textwidth]{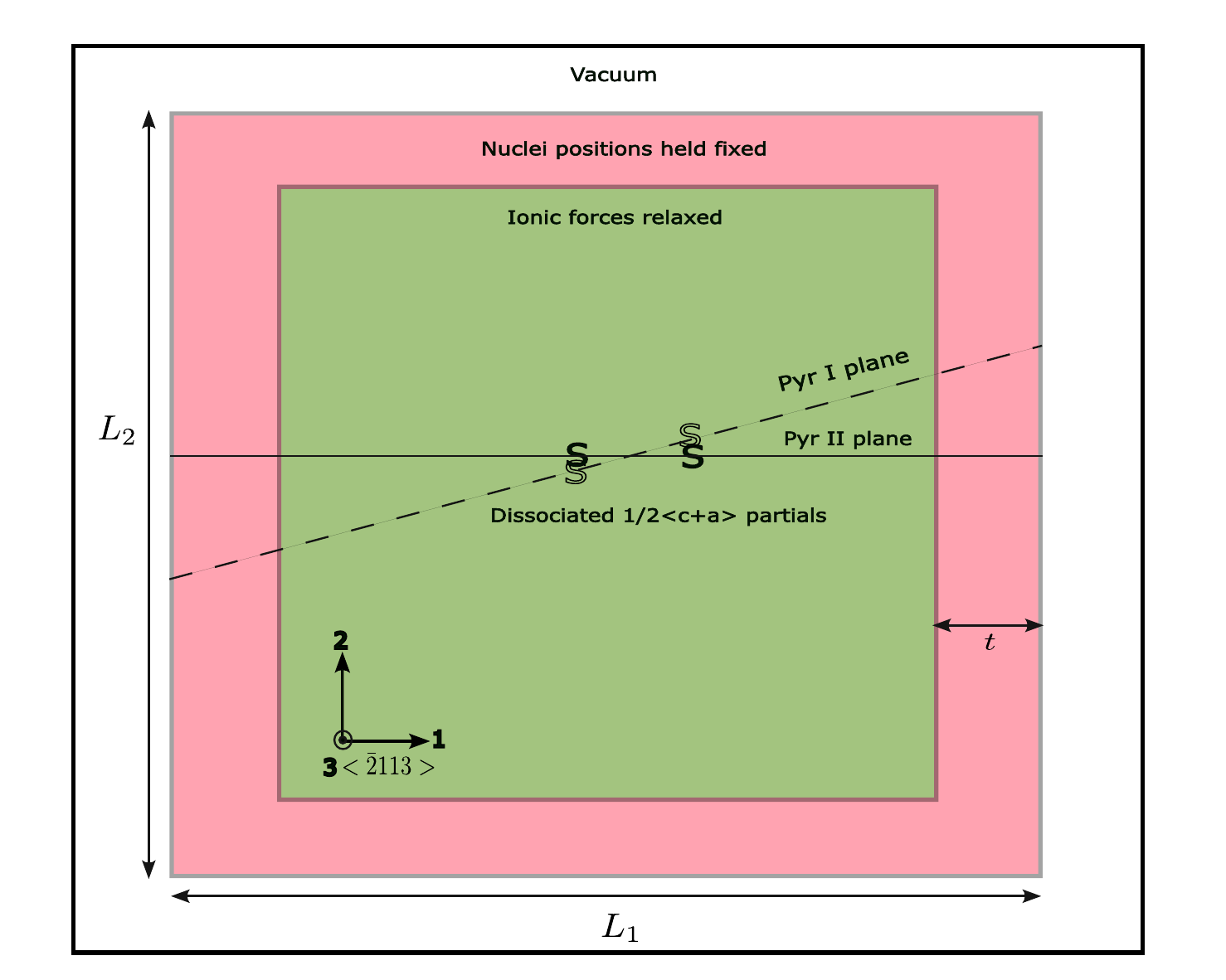}
  \caption{Schematic of simulation set up for dislocation core energetics calculations. Identical Volterra displacement boundary conditions corresponding to the perfect \cplusa ~screw dislocation is applied to both the Pyr II and Pyr I dissociated configurations in the boundary layer region of thickness $t$.}  
   \label{fig:coreEnergyCalcSchematic}
\end{figure}

We now detail the simulation set up to obtain the core energy difference between the pyramidal I and II \cplusa~screw dislocations in pure Mg (\deltaEpyr$=E^{\textrm {I}}_{\textrm{Mg}}-E^{\textrm{II}}_{\textrm{Mg}}$).  We aligned the 1-2-3 coordinate axes of the simulation domain along the crystallographic directions corresponding to the vector along the pyramidal II slip plane and normal to the \cplusa~dislocation line ($\hkl<-1010>$), normal vector to pyramidal II  $\hkl{2-1-12}$ slip plane , and along the \cplusa~dislocation line direction ($\hkl<-2113>$), respectively. As depicted in the schematic in Fig~\ref{fig:coreEnergyCalcSchematic}, we set up a cuboidal simulation domain containing a straight pyramidal \cplusa~I/II screw dislocation  at the center with semi-periodic boundary conditions. To elaborate, we applied non-periodic boundary conditions along  the 1-2 axes with a surrounding vacuum layer, and periodic boundary conditions along the axis 3. The simulation domain is thus composed of an atomic region and a vacuum region. The 1-2 dimensions of the atomic region are denoted by $L_1 \times L_2$, and the dimension along the periodic direction is the  $\frac{1}{3}\hkl<-2113>$ Burgers vector magnitude denoted by $b$. We also considered a dense shifted Monkhorst-Pack k-point grid of 12 k-points along the  periodic direction for sampling the Brillouin-zone, amounting to an effective supercell of size 12$b$ along the periodic direction for the electronic-structure calculation. Further, we divided the atomic region into an boundary layer region of thickness $t$ whose positions are held fixed and an inner dislocation core region which will undergo structural relaxation. In the boundary layer region, we used the classical Volterra solution of the perfect \cplusa~screw dislocation using anisotropic elasticity~\cite{Hirth1982} to apply the far field Dirichlet boundary condition on the atomic positions. The anisotropic elastic constants and lattice constants for HCP Mg were obtained from bulk DFT calculations and tabulated in Tab.~\ref{tab:elasticLatticeConstantsDFT}. In the inner dislocation core region, we introduced perturbations to the atomic positions to construct two different initial configurations corresponding to dissociation into $\frac{1}{2}$\cplusa~partials on the pyramidal I and II planes. In particular, we applied (a) a small partial separation of around 12 \AA~ along the pyramidal II plane, and (b) a similar small partial separation along the pyramidal I plane.  Subsequently, while applying homogeneous Dirichlet boundary conditions in 1 and 2 directions on the electronic fields, we performed ground-state pseudopotential Kohn-Sham DFT calculations along with structural relaxation of the ionic forces in the inner core region to under 5 meV/\AA. 
Finally, \deltaEpyr~is computed by taking the difference of the ground-state total energy of the pyramidal I and II systems. We note that \cplusa~I/II dissociated screw dislocations have the same far-field elastic displacement as they share the same Burgers vector. Thus  cancellation effects remove the spurious surface energy effects given a sufficiently large boundary layer thickness. In order to confirm this, we performed a convergence study of the ionic forces in the inner region with respect to the boundary layer thickness $t$, and find $t\sim$15 \AA~mitigates the surface effects on the ionic forces to be under 5 meV/\AA. Cell size convergence of \deltaEpyr~with respect to the non-periodic $L_1 \times L_2$ dimensions and sensitivity to DFT specific parameters will be discussed in Sec.~\ref{sec:resultDeltaEpyr}. We also note that these DFT calculations of  \deltaEpyr~are performed using a norm-conserving pseudopotential with 10 valence electrons per Mg atom that includes $3s^2$ and the semi-core $2p^6$ and $2s^2$ electrons. The requirement of a stringent accuracy pseudopotential is examined in Sec.~\ref{sec:resultsdftsensitivity}.

\begin{table}[htbp!]
\footnotesize
\begin{center}
\begin{tabular}{|c|c|c|c|c|c|c|c|c|}
\hline
a (\AA) & c (\AA) & $C_{11}=C_{22}$ (GPa) & $C_{33}$ (GPa) & $C_{44}$ (GPa)  & $C_{12}$ (GPa) & $C_{13}$ (GPa) & $C_{66}$ (GPa)\\
\hline
3.1967 & 5.1904 & 65.4 & 66.3 & 17.3 & 23.8 & 21.3 & 20.8 \\
 \hline
    \end{tabular}
\end{center}
    \caption{Computed pure Mg HCP relaxed lattice and transversely isotropic elastic constants obtained from DFT calculations using ONCV pseudopotential with 10 valence electrons per atom. 1-2 directions are in the basal plane. Elastic constants differ by up to 2 GPA with respect to experimental values at room temperature~\cite{MgElasticConstantsExpt}.}    
        \label{tab:elasticLatticeConstantsDFT}
\end{table}

\subsection{Computational methodology for \Us~ calculations}\label{sec:UsSetup}
We now discuss the computational method for computing the interaction energy between a \cplusa~pyramidal I/II straight screw dislocation and a single substitutional solute at a site $j$ near the dislocation core, denoted by \Us, from DFT calculations. The computation of \Us~for each pyramidal dislocation entails taking the difference of the total ground-state after structural relaxation of two dislocation plus solute systems: (system-A) one with the solute at the dislocation core site $j$ and another (system-B) with the solute far enough away (around 3 nm) from the dislocation-core center where the interaction between the dislocation stress field and the  misfit strain tensor of solute is negligible. Once we obtain \Us~for each pyramidal dislocation, we can compute \deltaUs, an important input to the cross-slip line tension model, by taking the difference of the \Us~values for pyramidal I and II dislocations: $\Delta U_j^{\textrm{I-II}}=U_j^{I}-U_j^{II}$. The simulation domain and the semi-periodic boundary conditions for the \Us~calculations are set up in the same manner as that for the \deltaEpyr~calculations in pure Mg (cf. Sec.~\ref{sec:deltaEpyrSetup}).  We reuse the relaxed \cplusa~core structures in pure Mg as starting structures. We consider a finite number of unique near core sites for solute substitution within a $b$ length along the \cplusa~dislocation line as the same sites are periodically repeated in each $b$ length. The near core sites are identified through  common neighbor analysis using the OVITO software (cf. Fig.~\ref{fig:CNA}) that marks the non-HCP core sites. Additionally, we include an outer ring of surrounding HCP sites, amounting to a total $N_s=58$ sites. Finally, the union of the above pyramidal I and II dislocation core sites are considered for solute substitution.  In Sec.~\ref{sec:resultsSoluteEffectsDFT}, we will analyze the  cell-size convergence of \deltaUs~with respect to the periodic length of the simulation box.


\section{Results and discussion}\label{sec:results}
\subsection{\deltaEpyr~from large-scale DFT calculations and convergence studies}\label{sec:resultDeltaEpyr}
Using the computational methodology detailed in Sec.~\ref{sec:deltaEpyrSetup}, we perform large-scale DFT calculations to compute the core energy difference between Pyr I and II  \cplusa~screw dislocations (\deltaEpyr). The calculations involve isolated straight screw dislocations in Mg employing semi-periodic boundary conditions (cf. Fig.~\ref{fig:coreEnergyCalcSchematic} for schematic of the setup).
Upon relaxation of the ionic forces, the dislocation cores split into $\frac{1}{2}$\cplusa~partials. However, the partials are not symmetric and have a non-planar stacking fault (cf. Fig~\ref{fig:panel1}(A)), as also observed in an earlier study~\cite{Itakura2016}.
As discussed in Sec.~\ref{sec:deltaEpyrSetup}, the elastic displacement field away from the dislocation core is the same for Pyr I and II \cplusa~screw dislocations. This allows computation of \deltaEpyr~from the difference of the total energy of the two systems with the spurious surface energy effects canceling out. Since \deltaEpyr~is a sensitive input to the cross-slip model, below we analyze the convergence of \deltaEpyr~with respect to the non-periodic cell-size ($L_1 \times L_2$) and DFT specific parameters.

\subsubsection{Cell-size convergence of \deltaEpyr}
\begin{figure}[htp]
 \centering
 \includegraphics[width=\textwidth]{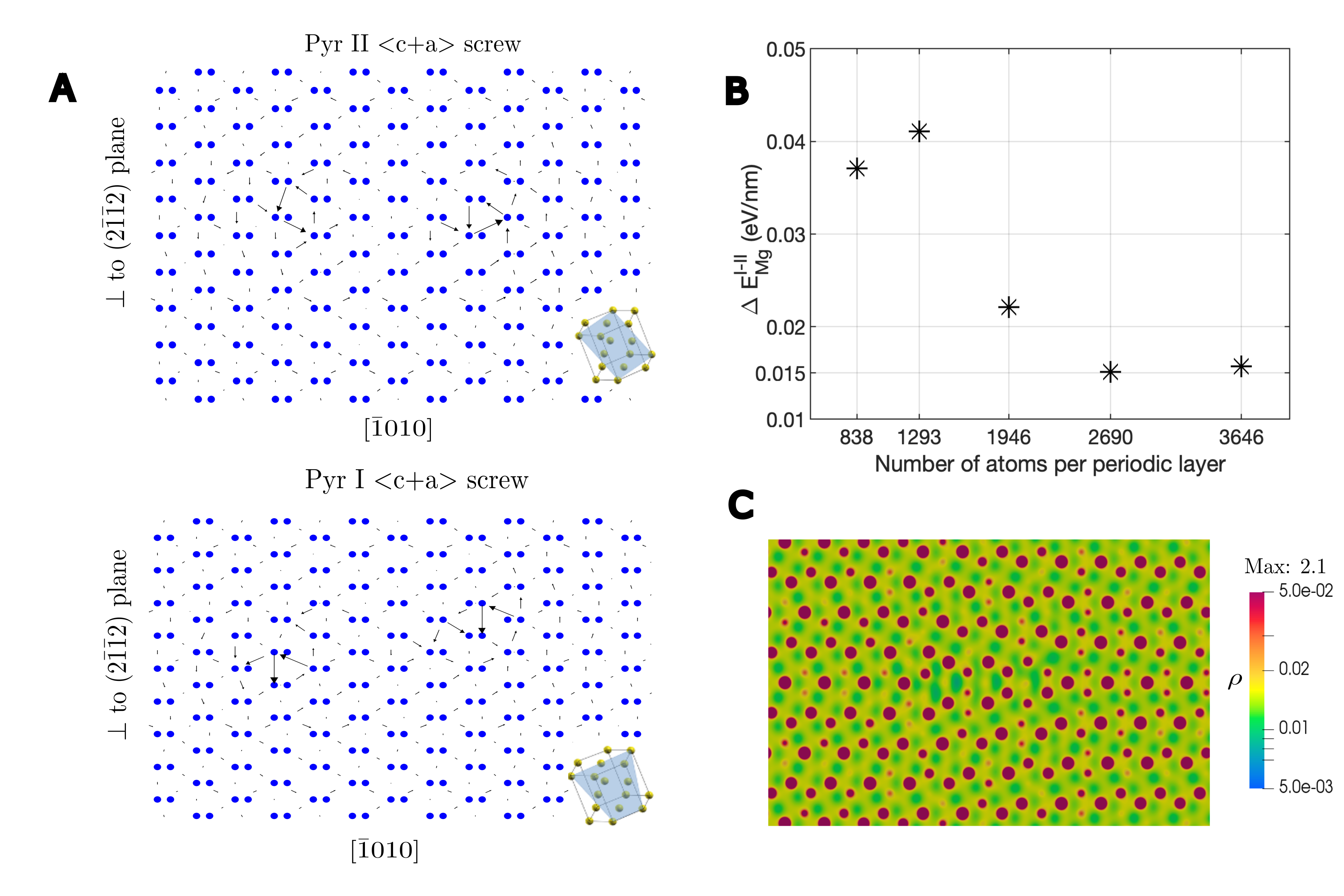}
    \caption{DFT calculations of Pyr I and II \cplusa~screw dislocations. (A) Differential displacement plots of relaxed dislocation cores showing non-planar nature (plane normal is $\hkl<-2113>$). (B) Cell size convergence of \deltaEpyr. (C) Electron density contour of Pyr II \cplusa~screw dislocation (plane normal is $\hkl<-2113>$).}  
 \label{fig:panel1}
\end{figure}

\begin{figure}[htpb]
 \centering
 \includegraphics[width=0.9\textwidth]{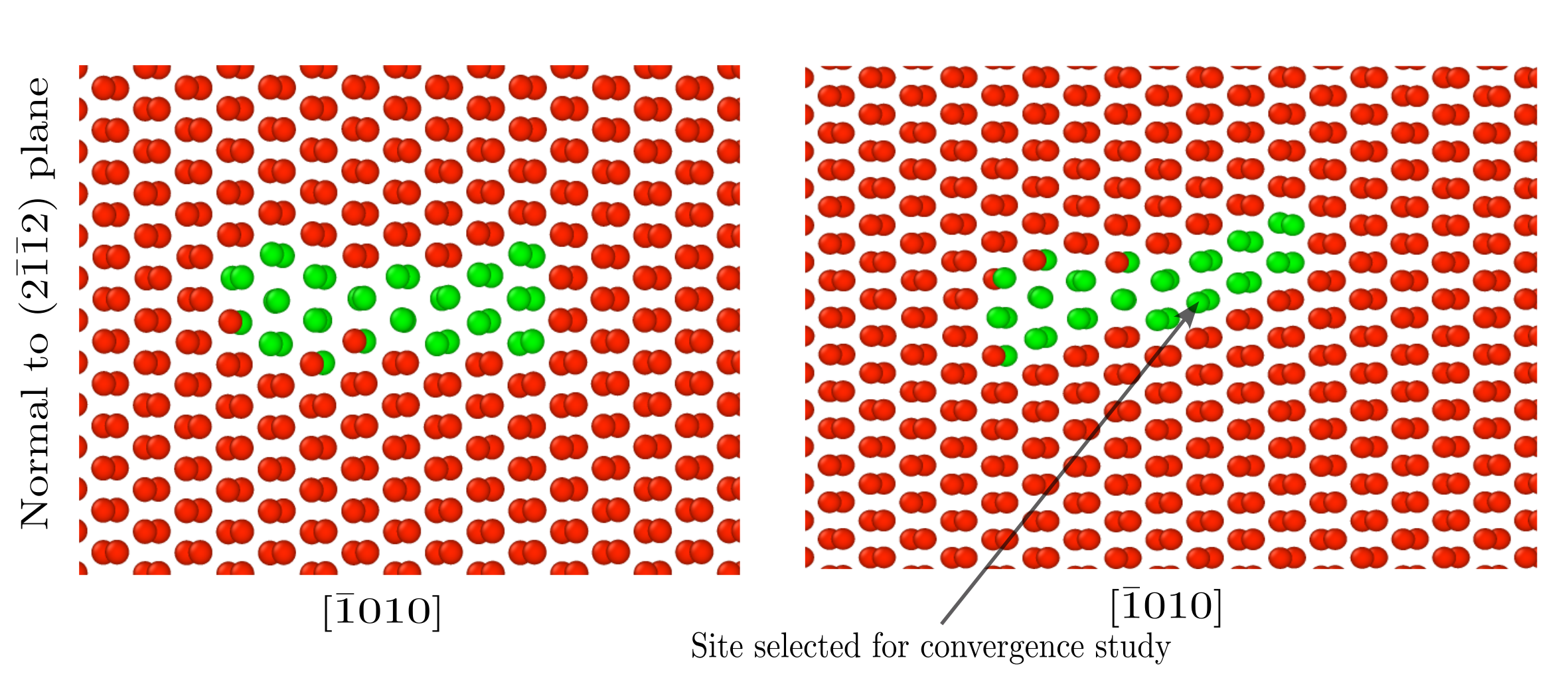}
  \caption{Common neighbor analysis of relaxed dislocation core structures. Left side is Pyr II  \cplusa~screw dislocation core structure and right side is Pyr I \cplusa~screw dislocation core structure. The solute site used for the cell-size convergence study of \deltaUs~in Tab.~\ref{tab:deltaUSensitivity} is also marked.} 
   \label{fig:CNA}
\end{figure}

We studied the non-periodic cell-size convergence of \deltaEpyr~by systematically increasing the $L_1 \times L_2$ dimensions from 6 nm $\times$ 5.3 nm  to 12.4 nm $\times$ 11.1 nm. In terms of number of atoms, this cell-size study considered simulation sizes of 838, 1293, 1946, 2690 and 3646 Mg atoms per periodic layer of length $b$ ($\frac{1}{3}\hkl<-2 1 1 3>$ Burgers vector magnitude) along the dislocation line. For each cell size,  the ionic forces in the inner core region (excluding the boundary layer of $\sim$15 \AA) are relaxed starting from the initial configuration chosen to be the relaxed structure corresponding to previous smaller cell size.   Table~\ref{tab:deltaECellsize} and Figure~\ref{fig:panel1}(B) show the results of the convergence study. Notably, we observe that a cell size of 2,700-3,600 atoms is required to obtain a converged value for \deltaEpyr~of 15 meV/nm.

%

\begin{table}[htp]
    \centering
    \small
    \begin{tabular}[t]{|c | c | c |}
        \hline
         Non-periodic dimensions                   &  Number of atoms (electrons) & \deltaEpyr      \\ 
            along 1-2 ($L_1 \times L_2$) (nm)                     & in simulation domain (per $b$) &   (meV/nm)  \\ \hline
         6.03 $\times$ 5.29  & 838 (8,380) & 37.1 \\
        7.62 $\times$ 6.35 & 1,293 (12,930) & 41.1\\
         9.21 $\times$ 7.94 & 1,946 (19,460) & 22.1\\
       10.80 $\times$ 9.53 & 2,690 (26,900)& 15.1\\
          12.38 $\times$ 11.11 & 3,646 (36,460) & 15.7\\
        \hline
    \end{tabular}
    \caption{Cell-size convergence analysis of \deltaEpyr.}    
        \label{tab:deltaECellsize}
\end{table}

\begin{table}[htp]
    \centering
    \small
    \begin{tabular}[t]{| c | c |}
        \hline
       Finite-element    & \deltaEpyr;~($7.62 \times 6.35$ nm); $1\times 1\times 6$ k-pts (shifted)    \\ 
              discretization (DoF/atom)                      &   (meV/nm)  \\  \hline 
        FEOrder=5, $h=0.61$ \AA~(15k)  &  48.77 \\  
        FEOrder=6, $h=0.61$ \AA~(26k)  &  37.79 \\
        FEOrder=6, $h=0.47$ \AA~(52k)    &  38.90\\
        \hline
    \end{tabular}\hfill%
    \begin{tabular}[t]{| c | c |}
        \hline
        Mg ONCV pseudopotential~\cite{van2018pseudodojo}                      & \deltaEpyr;~($10.80 \times 9.53$ nm);  $1\times 1\times 6$ k-pts (shifted)    \\ 
            valence electrons                         &   (meV/nm)  \\ \hline
        2 (3$s^2$)  &  25.9 \\
       10 (2$s^2$, 2$p^6$, 3$s^2$)  & 13.9 \\
        \hline
    \end{tabular}\hfill%
    \begin{tabular}[t]{| c | c |}
        \hline
        Monkhorst-Pack k-point                      & \deltaEpyr;~($10.80 \times 9.53$ nm)\\ 
                 grid                    &   (meV/nm)  \\      
        \hline
       $1\times 1\times 6$ (shifted)   & 13.9\\
       $1 \times 1\times 12$ (shifted) & 15.1 \\
        \hline
    \end{tabular}
    \caption{Sensitivity analysis of \deltaEpyr~with respect to \DFTFE parameters. The discretization and k-point sensitivity analysis was performed while keeping the relaxed dislocation core structure frozen. For the pseudopotential sensitivity study we relaxed the dislocation core structures for the 2  valence electrons ONCV pseudopotential starting from the relaxed structures obtained from 10 valence electrons ONCV pseudopotential. The non-periodic $(L_1 \times L_2)$ dimensions are indicated for each study.}    
        \label{tab:deltaESensitivity}
\end{table}

\subsubsection{Sensitivity analysis of \deltaEpyr~with respect to FE discretization,  pseudopotential, and k-points}\label{sec:resultsdftsensitivity}
Subsequently, we examined the sensitivity of \deltaEpyr~to other relevant parameters in \DFTFE that are the finite-element discretization, the  choice of the norm-conserving pseudopotential, and the k-point grid. First, we studied the convergence with respect to the finite-element discretization in DFT-FE. The finite-element discretization can be modified through both the polynomial degree of the Lagrange polynomial basis functions (FEOrder) and the finite-element size $h$ to obtain systematic convergence, finite-elements being a complete basis. Our convergence study in Tab.~\ref{tab:deltaESensitivity} demonstrates that $\textrm{FEOrder}=6$ and $h=0.47$ \AA~is sufficient to obtain a discretization accuracy of $\sim$ 1 meV/nm in \deltaEpyr. To provide context, the equivalent wavefunction cutoff energy parameter used in plane-wave DFT codes for a similar discretization accuracy would be around 100 Ha. Thus, all the DFT inputs (\deltaEpyr, \Us) in this work were calculated using $\textrm{FEOrder}=6$ and $h=0.47$ \AA. Next, we studied the influence of the choice of the Mg pseudopotential. Both these pseudopotentials are obtained from Pseudo-dojo database~\cite{van2018pseudodojo}. As shown in Tab.~\ref{tab:deltaESensitivity}, we observe a significant sensitivity of 12 meV/nm with respect to the number of Mg  electrons treated as valence in the ONCV pseudopotential, highlighting the importance of employing a stringent accuracy  ONCV pseudopotential that additionally includes the semi-core 2$s^2$, 2$p^6$ electrons as valence electrons along with the 3$s^2$ electrons. Prior studies of \deltaEpyr~\cite{Itakura2016,LIU2024119864} employed Mg pseduopotentials with 2 valence electrons (3$s^2$). Finally, we studied the sensitivity of \deltaEpyr~to the k-point grid density along the periodic line direction, as tabulated in Tab.~\ref{tab:deltaESensitivity}. We observe that  the difference between a dense ($1\times 1\times 12$)  and relatively coarser ($1\times 1\times 6$) Monkhorst-Pack grid to be 1.2 meV/nm. 

\subsection{Influence of external macroscopic deformation on \deltaEpyr}\label{sec:resultsMacroDeltaEpyr}
Next, we studied the influence of external non-glide macroscopic strains on \deltaEpyr. Such macroscopic strains (and stresses) are experienced by the pyramidal dislocations during the cross-slip process that is driven by external net resolved shear stress on the cross-slip plane. In particular, we considered the relaxed \cplusa~I/II screw dislocation core structures at the cell size consisting of 2,690 atoms and zero external strain, and  applied affine uniaxial strains along the 1, 2 and 3 directions ($\boldsymbol{\epsilon_{11}}$,  $\boldsymbol{\epsilon_{22}}$ and  $\boldsymbol{\epsilon_{33}}$), where 1-2-3 directions are oriented along $\hkl<-1010>$, normal to the Pyr II $\hkl{2-1-12}$ plane  and along the \cplusa~direction ($\hkl<-2113>$). Subsequently, we performed ground-state DFT calculations to compute \deltaEpyrmacro. We applied strain values of $\pm 0.2\%$ for each of the uniaxial strains. The values of the \deltaEpyrmacro~at the different strain values are tabulated in Tab.~\ref{tab:stressDependency}.  They demonstrate a strong influence of \deltaEpyr~on external strain, particularly for the $\boldsymbol{\epsilon_{11}}$,   $\boldsymbol{\epsilon_{22}}$ strains. This has an important consequence on the barrier for cross-slip as will be discussed subsequently. We also note that \deltaEpyrmacro~has an asymmetric dependency with respect to $\boldsymbol{\epsilon_{22}}$ strain (normal to Pyr II plane).
\begin{table}[htbp!]
\begin{center}
\begin{tabular}{||c|c|c||}
\hline
Uniaxial non-glide strains &\deltaEpyr~($+ 0.2\% \epsilon$)  &\deltaEpyr~($- 0.2\% \epsilon$)     \\
 & (meV/nm) & (meV/nm)   \\
\hline
 $\boldsymbol{\epsilon_{11}}$ & 13.5 & 19.3   \\
\hline 
$\boldsymbol{\epsilon_{22}}$ & 1.3 & 15.3   \\
 \hline 
$\boldsymbol{\epsilon_{33}}$ & 15.0 & 13.8   \\
 \hline\hline
    \end{tabular}
\end{center}
    \caption{Influence of non-glide strains on \deltaEpyr~for a system size containing 2690 atoms per periodic layer. The zero strain \deltaEpyr~value is 15.1 meV/nm. The 1-2-3 directions are aligned along $\hkl<-1010>$, normal to the pyramidal II $\hkl{2 -1 -1 2}$ plane  and along the \cplusa~direction ($\hkl<-2 1 1 3>$).}    
        \label{tab:stressDependency}
\end{table}

\subsection{Solute effects on \deltaEpyr~from large-scale DFT calculations}\label{sec:resultsSoluteEffectsDFT}
\begin{figure}[htp]
 \centering
 \includegraphics[width=0.9\textwidth]{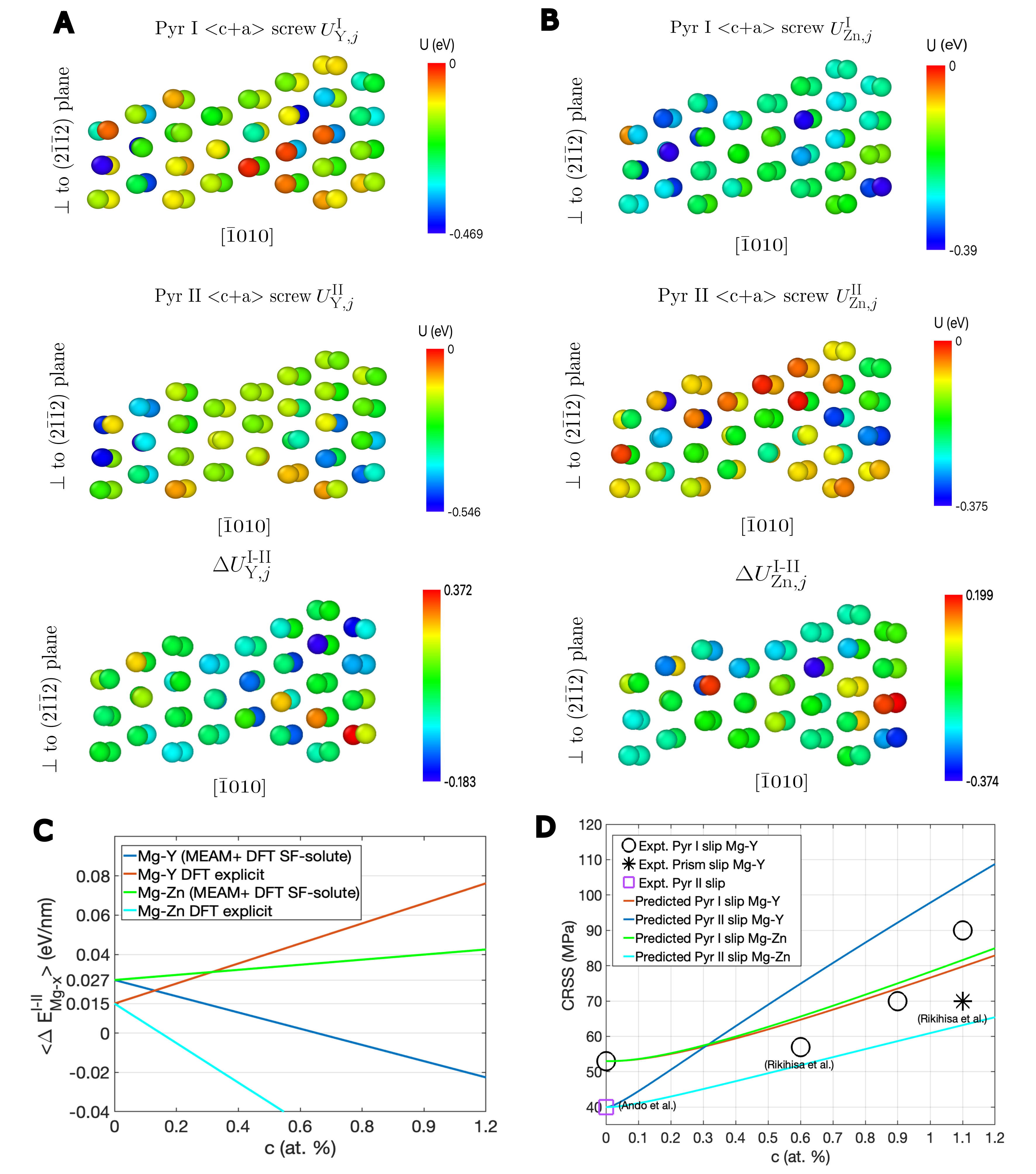}
   \caption{(A) and (B) Pyr I and II \cplusa~screw dislocation Y and Zn solute interaction energies (\Us) at various near-core sites. $\Delta U_j^{\textrm{I-II}}$ values for Y and Zn are also plotted on the Pyr I relaxed core structure. The sites were identified using common neighbor analysis and we additionally included an outer ring of surrounding HCP sites, amounting to a total $N_s=58$ sites.  (C) average solute effects on \deltaEpyr~in Mg-Y and Mg-Zn, comparison between explicit DFT \deltaUs~inputs and previous MEAM/DFT SF-solute based approach~\cite{WuCurtinScience2018,AhmadCurtin2019} (D) concentration dependent critical resolved shear stress (CRSS) values predicted using a Labusch type weak pinning model~\cite{leyson2010quantitative} and the computed \Us~inputs for Y and Zn solutes. The reference CRSS values for FPCS slip (Pyr I) in single crystal Mg-Y are taken from Ref.~\cite{Rikihisa2017}. Pure Mg single crystal CRSS values at room temperature are approximately 40 MPa and 53 MPa for SPCS and FPCS glide respectively~\cite{ando2010deformation}.}  
 \label{fig:panel2}
\end{figure}

We computed the dislocation solute interaction energies (\Us) of \cplusa~pyramidal I/II screw dislocations, using direct DFT calculations, for various locations of the solute in the dislocation core. The computational methodology of the \Us~calculations and details regarding the sampling of the core sites is discussed in Sec.~\ref{sec:UsSetup}. We now analyze the cell-size convergence of $\Delta U_j^{\textrm{I-II}}=U_j^{I}-U_j^{II}$, where the primary sensitivity stems from the spurious solute-solute interactions due to the periodic boundary condition along the dislocation line. In Tab.~\ref{tab:deltaUSensitivity}, we examined the  convergence of \deltaUs~for a chosen core site with respect to  the periodic cell dimension. We find that a periodic length of at-least $4b$ is required to reduce the sensitivity in \deltaUs~to below 10 meV. Accordingly, for all the \Us~calculations, we choose the periodic dimension to be 4$b$ and use a Gamma point for sampling the Brillouin-zone.  
Furthermore, we have used the softer ONCV pseudopotential with 2 valence electrons ($3s^2$) due to the relatively less stringent accuracy required for the \deltaUs~compared to quantifying \deltaEpyr, that has a much smaller value.  
We have considered two ductility favorable solutes in this work: Y and Zn. Figures~\ref{fig:panel2} (A-B) shows the map of \Us~values at various near core sites sampled from the union of pyramidal I and II dislocation cores. The corresponding \deltaUs~values for Y and Zn are also shown in Fig.~\ref{fig:panel2} (A-B). 
\begin{table}[ht]
    \centering
    \small
    \begin{tabular}[t]{|c| c | c | c |}
        \hline
        Periodic   & Non-periodic                    &  Number of atoms (electrons)& $\Delta U_{\textrm{Y},j}^{\textrm{I-II}}$      \\ 
           length    & dimensions along 1-2  (nm)                     & in simulation domain  &   (meV)  \\ \hline
        2$b$  & 8.15 $\times$ 6.88  & 3,008 (6,025) & 125  \\
        3$b$  & 8.15 $\times$ 6.88 & 4,512 (9,033) & 192 \\
        4$b$  & 8.15 $\times$ 6.88 & 6,016 (12,041) & 128\\
        5$b$  & 8.15 $\times$ 6.88 & 7,520 (15,049) & 133\\        
        \hline
    \end{tabular}
    \caption{Cell-size convergence analysis of $\Delta U_{\textrm{Y},j}^{\textrm{I-II}}$ for a chosen near-core solute site (cf. site marked in Fig.~\ref{fig:CNA}) with respect to length of the simulation cell along the periodic direction (dislocation line). For each periodic length choice we have converged the $\Delta U_{\textrm{Y},j}^{\textrm{I-II}}$ value with respect to the Monkhorst-Pack k-point grid.}    
        \label{tab:deltaUSensitivity}
\end{table}

Using the ab-initio computed \deltaEpyr~and \Us~values (for Y and Zn solutes) from direct large-scale DFT calculations, we discuss the estimation of cross-slip barrier and implications to ductility with increasing solute concentration in Mg-Y and Mg-Zn alloys.

\subsection{Predictions using DFT computed \deltaEpyr~and solute effects on \deltaEpyr} 
In dilute Mg alloys, the cross-slip from the lower energy Pyr plane to the higher energy Pyr  plane is the rate limiting step in the double cross-slip mechanism proposed in Refs.~\cite{WuCurtinScience2018,AhmadCurtin2019}.
The associated line-tension model for estimating the cross-slip energy barrier requires \deltaEpyr~and \deltaUs~as inputs, where \deltaUs~is the site based dislocation-solute interaction energy difference between the Pyr I and Pyr II screw dislocations ($\Delta U_j^{\textrm{I-II}}=U_j^{I}-U_j^{II}$). The $\Delta U_j^{\textrm{I-II}}$ values are used to estimate the change in \deltaEpyr~in a dilute random alloy, using statistical analysis, which is denoted by $\Delta E^{\textrm{I-II}}_{\textrm{Mg-x}}$ where $\rm{x}$ denotes the solute type. $\Delta E^{\textrm{I-II}}_{\textrm{Mg-x}}$ accounts for two effects: (i) the average solute effect  $\langle\Delta E^{\textrm{I-II}}_{\textrm{Mg-x}}\rangle$ that scales linearly with the solute concentration ($c$); (ii) the fluctuation solute effect, denoted by $\mathcal{F}\left(\Delta E^{\textrm{I-II}}_{\textrm{Mg-x}}\right)$, which reduces the energy barrier by availing favorable solute sites. The fluctuation contribution scales as $\sqrt{c}$. We refer to ~\ref{sec:appCrossSlipModel} for an outline of the line-tension model.

We now discuss the significant differences in the findings using direct DFT calculations of \deltaEpyr~and~\Us~obtained in this work, compared to the findings in Ref.~\cite{WuCurtinScience2018} where MEAM interatomic potentials were employed to compute \deltaEpyr~and a surrogate approach, using DFT computed stacking fault solute interaction energies with elastic corrections, was used to estimate \Us. We consider the binary alloys Mg-Y and Mg-Zn as the two prototypical cases, where ductility enhancement over pure Mg is observed in experiments. 
We revisit the central reasoning put forth for intrinsic ductility enhancement in certain Mg-RE alloys, which was attributed to a reduction of \deltaEpyr~in the presence of some solute types resulting from  negative values in the slope of $\langle\Delta E^{\textrm{I-II}}_{\textrm{Mg-x}}\rangle$ with respect to solute concentration. In the case of Y solute, experiments suggest a large ductility enhancement in Mg with 0.5-0.9 at.\% Y~\cite{Rikihisa2017,Rikihisa2020}. The \Us~values obtained via the surrogate approach suggested $\langle\Delta E^{\textrm{I-II}}_{\textrm{Mg-x}}\rangle$ to decrease with increasing concentration, with a slope of -41.3~meV/nm/at\%, thus seemingly providing an explanation for the enhancement of ductility observed experimentally. On the contrary, direct DFT calculations of \Us~provide an increasing $\langle\Delta E^{\textrm{I-II}}_{\textrm{Mg-x}}\rangle$ with concentration, with a slope of 51~meV/nm/at\% (cf. Fig.~\ref{fig:panel2} (C)). The resulting cross-slip energy barrier increases with increasing Y concentration, suggesting that relative stabilization of the energy of Pyr~I screw dislocation with respect to Pyr~II may not be the central reason for enhanced intrinsic ductility. The $\langle\Delta E^{\textrm{I-II}}_{\textrm{Mg-x}}\rangle$ trends are also opposing in the case of Zn solutes (cf. Fig.~\ref{fig:panel2} (C)), which suggests that the surrogate approach of using stacking fault solute interaction energies may not provide even a qualitatively correct picture.

\subsection{Incorporation of solute-strengthening and effect of macroscopic deformation on core energetics}\label{sec:incorporation}
\begin{figure}[htp]
 \centering
 \includegraphics[width=\textwidth]{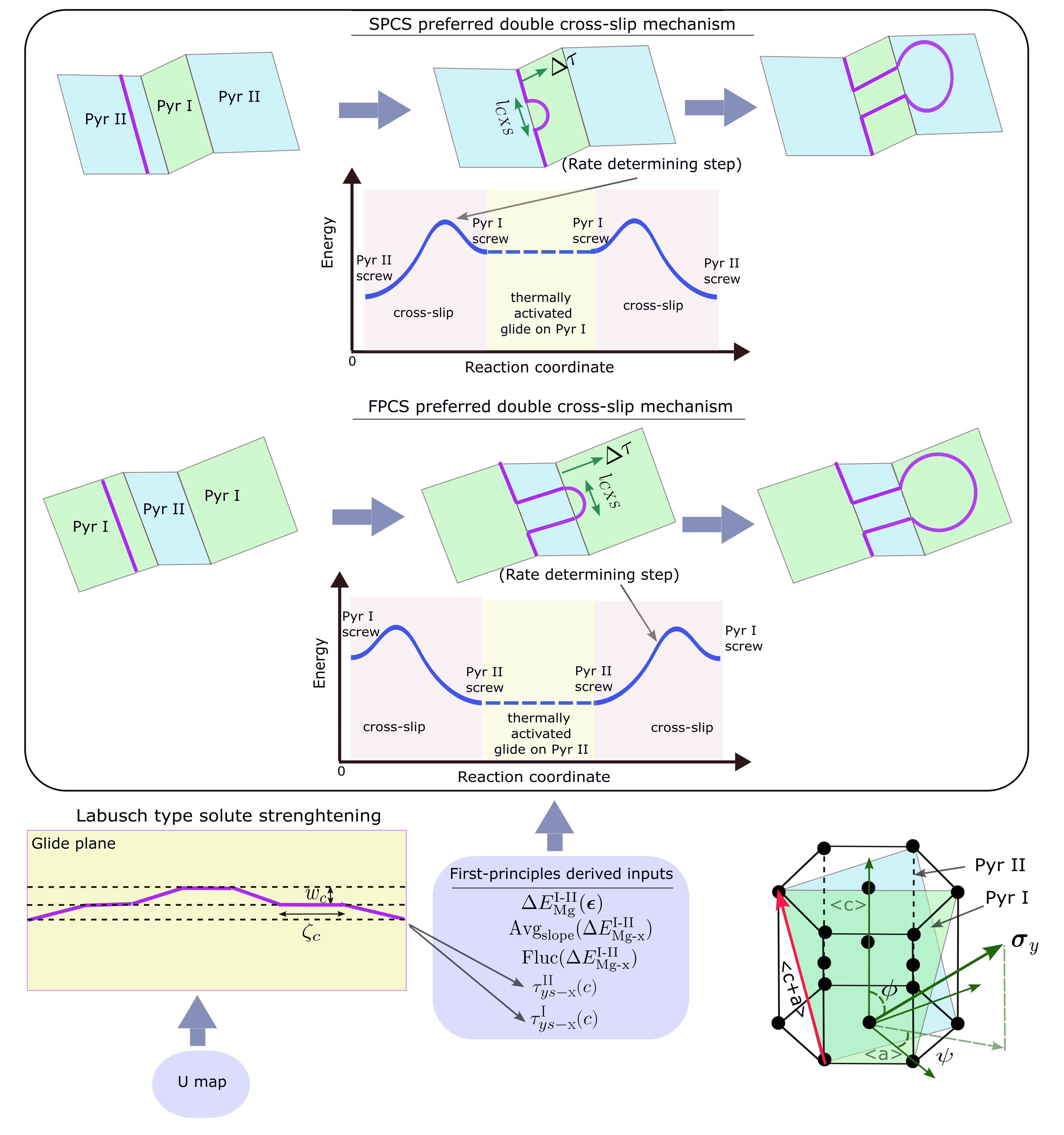}
\caption{Schematic of the  double cross-slip mechanism in dilute Mg-Y alloy depicting the different slip preference regimes (SPCS/FPCS). The first-principles dislocation and dislocation-solute energetics inputs, and first-principles derived solute-strengthening inputs to the cross-slip energy barrier model are indicated.} 
   \label{fig:lineTensionModelSchematicMgY}
\end{figure}

\begin{figure}[htp]
 \centering
 \includegraphics[width=\textwidth]{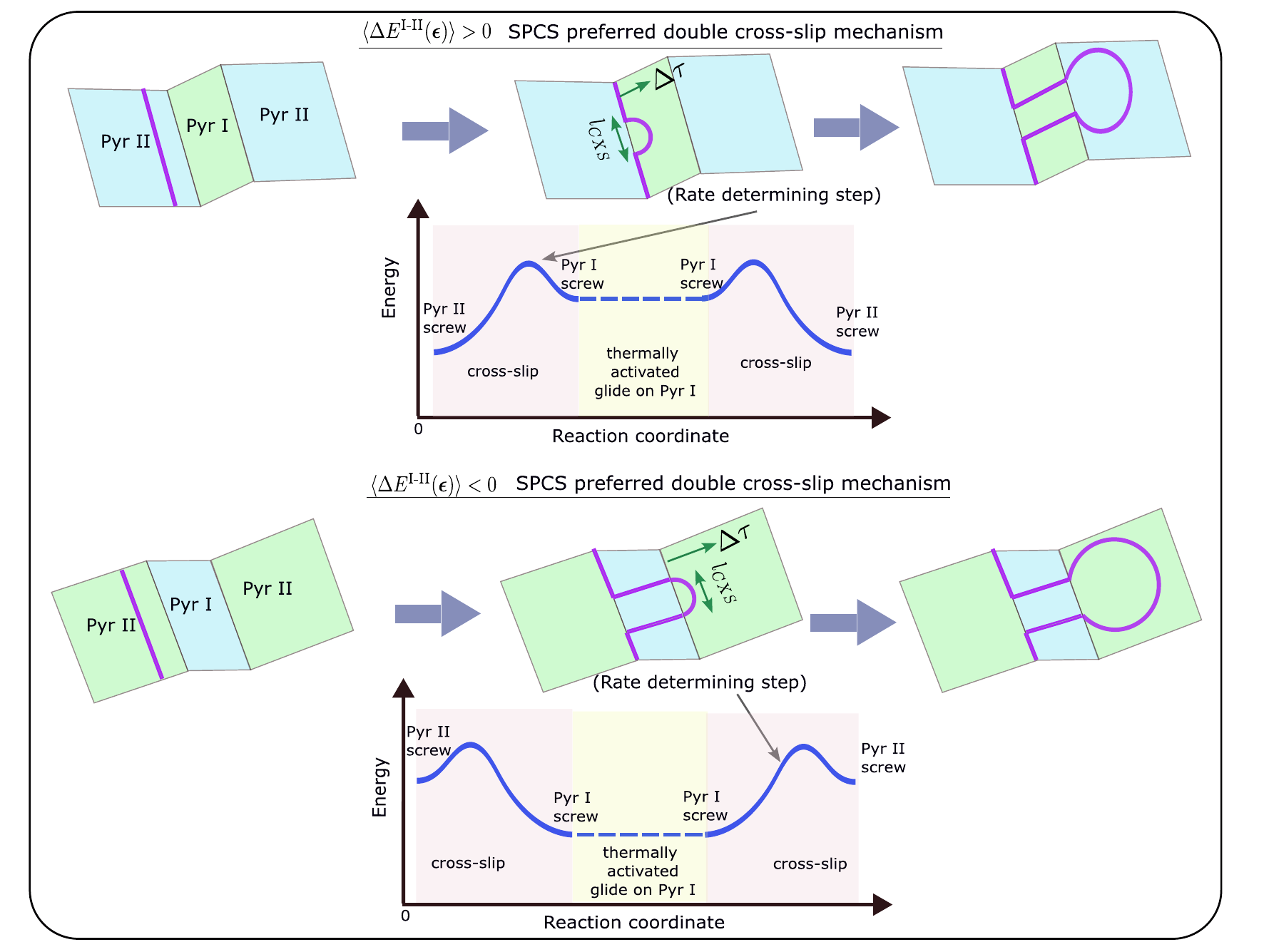}
  \caption{Schematic of the  double cross-slip mechanism in dilute Mg-Zn. SPCS is predicted to the preferred slip mode across all dilute concentrations.  The cross-slip from the lower energy Pyr II to the Pyr I plane is the rate determining step when $\langle\Delta E^{\textrm{I-II}}_{\textrm{Mg-x}}(\boldsymbol{\epsilon})\rangle >0$ and the reverse for $\langle\Delta E^{\textrm{I-II}}_{\textrm{Mg-x}}(\boldsymbol{\epsilon})\rangle <0$.}  
   \label{fig:lineTensionModelSchematicMgZn}
\end{figure}

Sustained cross-slip of Pyr dislocation in Mg is possible via  double cross-slip of Pyr screw dislocations involving triplets of intersecting Pyr I-Pyr II planes~\cite{WuCurtinScience2018,AhmadCurtin2019,AhmadCurtin2020}. The energy barrier of the rate determining step in the double cross-slip process is denoted by $\Delta G_{XS}(c)$.
A key aspect not considered in the aforementioned dislocation line tension model to quantify $\Delta G_{XS}(c)$ is the solute strengthening effects on Pyr screw dislocation glide. As will be discussed below, incorporation of solute strengthening effects based on first-principles \Us~input  significantly influences the primary slip system and the external driving force for the cross-slip mechanism. The other key consideration is the dependency of $\Delta E^{\textrm{I-II}}_{\textrm{Mg}}$ on non-glide macroscopic strain/stress, that we compute from our first-principles calculations ($\Delta E^{\textrm{I-II}}_{\textrm{Mg}}(\boldsymbol{\epsilon})$ as tabulated in Tab.~\ref{tab:stressDependency}). We  demonstrate that the  macroscopic strain dependence also  substantially influences $\Delta G_{XS}(c)$. Below, we detail how we incorporate these aspects into the  cross-slip  model, building upon earlier work~\cite{WuCurtinScience2018}. To aid our discussion, we schematically depict the  double cross-slip process for Mg-Y alloy in Fig~\ref{fig:lineTensionModelSchematicMgY}, and indicate the first-principles based inputs to the model.

Solute strengthening results in critical resolved shear stress (CRSS) values for Pyr I and II screw dislocation being higher than that of pure Mg. We use the DFT computed dislocation-solute interaction energies (\Us) in a Labusch type weak pinning model~\cite{leyson2010quantitative} to estimate the solute concentration dependent CRSS at room temperature 
(cf. ~\ref{sec:appCRSS}). The CRSS value, denoted by $\tau_{CRSS}(c)$, is the sum of the following two contributions: the CRSS in pure Mg denoted by $\tau_{y0}$; and the solute mediated strengthening  denoted by  $\tau_{ys-\textrm{x}}(c)$. Figure~\ref{fig:panel2}(D) shows that $\tau_{CRSS}$ for Pyr II screw dislocation increases at a faster rate with respect to Y concentration compared to that for Pyr I screw dislocation. The computed $\tau_{CRSS}$ for Pyr I screw dislocation is in good agreement with experimental values in Ref.~\cite{Rikihisa2017}. The results also suggest Pyr II $\tau_{CRSS}$ exceeds Pyr I $\tau_{CRSS}$ beyond  0.3  at.\%  Y. On the contrary, in case of Mg-Zn there is no such cross-over between Pyr I and Pyr II $\tau_{CRSS}$, with Pyr I $\tau_{CRSS}$ being consistently higher by about 15 MPa over Pyr II $\tau_{CRSS}$ in Mg-Zn across the dilute concentration range. As will be discussed below, this has a key bearing on the qualitatively different primary slip system preference observed in Mg-Y and Mg-Zn as a function of solute concentration, where the primary Pyr slip system (SPCS/FPCS) indicates the one where the dislocation glide is relatively more favored. The primary slip system is determined using our computed concentration dependent $\tau_{CRSS}$ in a local quasi-static equilibrium condition on the Pyr screw dislocations.

We now discuss the above solute strengthening mediated effects. The primary slip system is determined by comparing the local quasi-static equilibrium yield stress state $\boldsymbol{\sigma}_y$ of the Pyr I and II \cplusa~screw dislocations. The stress state $\boldsymbol{\sigma}_y$ is created by the external tensile loading on the sample along a loading direction characterized by the polar $\phi$ and azimuth $\psi$ angles (cf. Fig.~\ref{fig:lineTensionModelSchematicMgY}). We take our $\phi$ and $\psi$ values commensurate with tensile test experiments for the alloy in question (cf. Tabs. ~\ref{tab:modelInputsPureMg}--~\ref{tab:modelInputsMgZn}). We estimate the local quasi-static equilibrium  yield stress magnitude $\sigma_y(c)$ along the loading direction as the minimum between the two Pyr planes:
\begin{align}\label{eq:quasistaticEqCondition}
   \sigma_y(c) = &\textrm{min}\{\frac{\tau^{II}_{CRSS}(c)+\tau_b}{m_{II}(\phi)},\,\,\frac{\tau^{I}_{CRSS}(c)+\tau_b}{m_{I}(\phi)}\}\,,
\end{align}
where $m_{II}(\phi)$ and $m_{I}(\phi)$ are the  Schmidt factors on Pyr II and Pyr I plane, respectively, and $\tau_b$ signifies an additional local back stress beyond the single crystal CRSS that is set to a fixed value for all concentrations and solute types (Y and Zn).  We motivate the inclusion of $\tau_b$ from experimentally observed strain hardening effects in room-temperature tensile tests  in both single and poly-crystal  Mg-Y and Mg-Zn alloys~\cite{Rikihisa2017,Rikihisa2020,ando2012plasticMgZn,shi2013effects}. These strain hardening effects can be caused by dislocation pileups and grain-boundaries in the poly-crystalline case. These effects can increase the quasi-static yield stress magnitude. We note that such additional strengthening effects were taken into account in earlier works to determine the yield stress in the dislocation line tension model~\cite{WuCurtinScience2018,AhmadCurtin2019}. 
In this work, we use a value of $\tau_b= 12$ MPa. We infer this value from the  strain hardening magnitude in the stress-strain curve of single and poly-crystal Mg-Y around onset of yielding  (cf. Fig. 3 in Ref.~\cite{Rikihisa2017} and Fig. 3 in Ref.~\cite{Rikihisa2020}). Applying the above quasi-static equilibrium condition, the preferred Pyr plane (Pyr II or Pyr I, or equivalently SPCS/FPCS) for a given $c$ value is decided based on which provides a lower $\sigma_y(c)$ value.  In Mg-Y, this leads to  different slip preference regimes, which is SPCS at lower concentrations (till 0.5 at.\% Y) and transitions to FPCS at moderate and higher concentrations. The different primary slip preference  regimes and their associated double-cross slip mechanisms, as will be discussed below, is illustrated in Fig.~\ref{fig:lineTensionModelSchematicMgY}. In the case of Mg-Zn, the quasi-static equilibrium condition indicates only SPCS slip preference with no transition to FPCS preference (cf. schematic Fig.~\ref{fig:lineTensionModelSchematicMgZn}).

With the preferred slip system determined from $\tau_{CRSS}^{I/II}(c)$ and Schmidt factors, we sketch out the double-cross slip process involving the following three steps: 1) cross-slip from preferred to non-preferred Pyr slip plane; 2) constant strain-rate glide or a thermally activated glide process on the non-preferred Pyr slip plane which has the higher yield stress value; and 3) cross-slip from non-preferred Pyr slip plane to an adjacent preferred Pyr slip plane thereby facilitating dislocation multiplication through further double cross-slip events. The above steps are also illustrated in Fig.~\ref{fig:lineTensionModelSchematicMgY}. 
Among the two cross-slip events above, we consider the rate determining step to be the cross-slip from the lower (``$l$'') energy Pyr plane to the higher energy (``$h$'') plane (can either be step 1 or 3), whose barrier we are interested in quantifying. We remark that the cross-slip from the higher to lower energy Pyr plane would have a much smaller barrier, incurring  only the nucleation step barrier in the line-tension model (cf. ~\ref{sec:appCrossSlipModel}). The rate determining cross slip step is driven by the net resolved shear stress $(\Delta \tau_{RSS}^h (c)=\tau_{RSS}^{h}(c)-\tau_{CRSS}^h(c))$ on the higher energy Pyr plane, where $\tau_{RSS}^{h}(c)$, the net resolved shear stress, is  generated by the external tensile loading $\sigma_y(c)$ (cf. Eq~\ref{eq:quasistaticEqCondition}). $\Delta \tau_{RSS}^h(c)$ has to be positive to achieve sustained cross-slip that involves bowing out of the dislocation line on the cross-slip plane. Subsequently, we consider the intervening thermally activated  glide step on the non-preferred Pyr plane between the two cross-slip steps. Here the resolved shear stress, $\tau_{RSS}^{np}(c)$ (``$np$'' denotes non-preferred), if greater than the concentration dependent $\tau_{CRSS}^{np}(c)$ can enable a constant strain-rate glide on the non-preferred Pyr plane. On the contrary, if  $\tau_{RSS}^{np}(c)$ is smaller than the solute strengthened $\tau_{CRSS}^{np}(c)$ but larger than the pure crystal CRSS value ($\tau_{y0}^{np}$), it can only enable thermally activated glide of the weak pinned  dislocation line in a random field of solutes. Further, we assume no glide process is possible if $\tau_{RSS}^{np}(c)$ is smaller than the pure crystal Pyr slip CRSS value\footnote{It is reasonable to not consider room temperature thermal activation over pure Mg Pyr slip barriers (40-60 MPa CRSS).}. These unfavorable conditions for glide can arise when there is a larger gap between the Pyr I and Pyr II $\tau_{CRSS}$.  Thus, for a given solute concentration value, the following two  conditions are required for  double cross-slip to be possible:  
  \begin{subequations}
\begin{align}
  \textrm{Necessary condition for cross-slip: } 
  \Delta \tau_{RSS}^h (c)  &> 0 
  \label{eq:crossSlipCondition} \\
  \textrm{Necessary condition for thermally activated glide: } 
  \tau_{RSS}^{np}(c) &> \tau_{y0}^{np} \label{eq:thermallyActivatedGlide} 
\end{align}
\end{subequations}
In order to assess the consequences of the above conditions, we consider  Mg-Y and Mg-Zn single crystals under tensile loading along $\hkl<1 1 -2 0>$, and investigate the various concentration regimes. First, in the case of Mg-Y, we find that the necessary condition for cross-slip $(\Delta \tau_{RSS}^h (c)  > 0)$ is not satisfied at very small Y concentrations less than 0.17 at.\% Y, and thus double cross-slip mechanism is not likely activated in this regime. This condition, however, is satisfied for Y concentrations greater than 0.17 at.\% Y. Considering the second condition on the thermally activated glide, we find that this is always satisfied at concentrations greater than  0.12 at.\% Y. In the second case pertaining to Mg-Zn single-crystal, we find that below 0.3 at.\% Zn, the resolved shear stress on the non-preferred Pyr I plane is sufficiently high to satisfy the necessary condition for thermally activated glide. On the other hand, for Zn concentration above $\sim$0.05 at.\%, the necessary condition for cross slip (Eq.~\ref{eq:crossSlipCondition}) is satisfied.

\begin{figure}[htp]
 \centering
\includegraphics[width=0.8\textwidth]{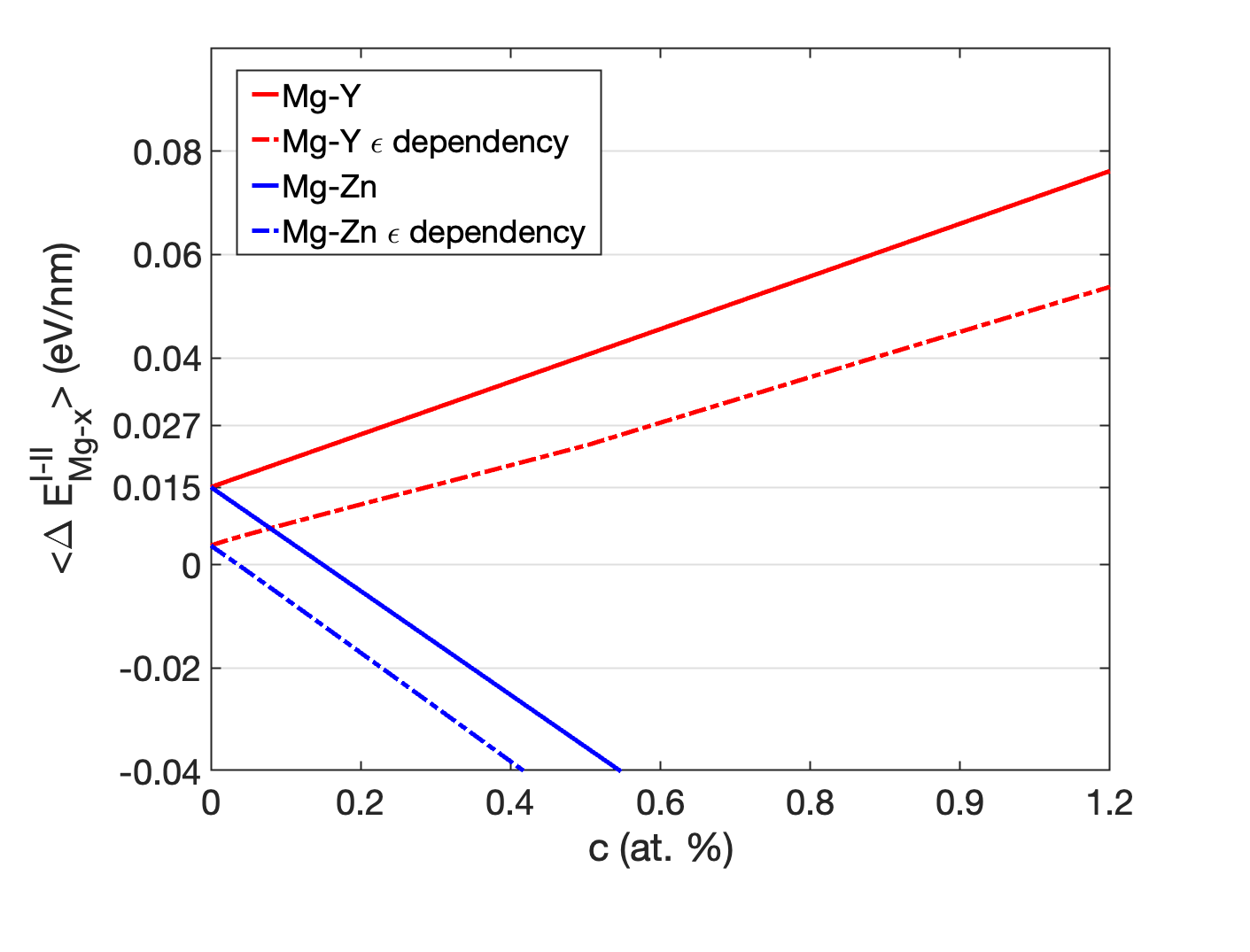}
  \caption{DFT computed  average solute effects on \deltaEpyr~in Mg-Y and Mg-Zn single crystal, with and without non-glide external macroscopic stress dependency.}  
   \label{fig:deltaEMacroSolute}
\end{figure}

We next present the importance of incorporating the dependence of $\Delta E^{\textrm{I-II}}_{\textrm{Mg}}$ on non-glide macroscopic strain/stress. The macroscopic stress state, $\boldsymbol{\sigma}_y(c)$ (cf. Eq.~\ref{eq:quasistaticEqCondition}), stems from the local quasi-static condition, as discussed before. We model this dependence based on first-principles calculations of \deltaEpyr~subjected to affine uniaxial strains (cf. Tab.~\ref{tab:stressDependency}). We approximate the strain dependence to be linear about each strain component, except for $\boldsymbol{\epsilon_{22}}$ (normal to Pyr II plane), where we consider a piecewise linear dependence due to the strong tension/compression asymmetric influence on \deltaEpyr. We refer to ~\ref{sec:appMacro} for more details on evaluation of $\Delta E^{\textrm{I-II}}_{\textrm{Mg}}(\boldsymbol{\epsilon})$ including the determination of the strain tensor $\boldsymbol{\epsilon}$ from $\boldsymbol{\sigma}_y(c)$.  In Fig.~\ref{fig:deltaEMacroSolute}, we plot \deltaEpyrmacro~as a function of solute concentration for Mg-Y and Mg-Zn single crystals under tensile loading. We observe that the value of \deltaEpyr~is reduced by 10--20 meV/nm across the concentration range, which is a substantial reduction.  
As will be demonstrated in the subsequent section, this reduction in \deltaEpyr~has a significant influence in reducing the cross-slip energy barrier in Mg-Y. 
Additionally, we remark that since $\boldsymbol{\sigma}_y$ is dependent on the concentration, $\boldsymbol{\epsilon}$ and thereby \deltaEpyrmacro~ also are solute concentration dependent.  

We incorporate the above discussed solute strengthening and macroscopic deformation effects into a dislocation line tension model (cf. ~\ref{sec:appCrossSlipModel})  for predicting the cross-slip energy barrier. The details of their incorporation and the expression of the line tension model are discussed therein.

\subsection{Cross-slip energy barriers and ductility enhancement predictions for Mg-Y}
\begin{table}[htbp!]
\footnotesize
\begin{center}
 \begin{tabular}{|c|c|c|}
\hline
Input & Source/Remark &  Value \\
\hline
\hline
 \deltaEpyr~(meV/nm) & DFT computed  & 15 \\
 \hline
 \deltaEpyr~non-glide uni-axial strain dependency  & DFT computed & Tab.~\ref{tab:stressDependency} \\
 \hline
 $\tau^{\textrm{II}}_{y0}$ (MPa) & Experimental 300 K~\cite{ando2010deformation}  & 40 \\ 
 \hline
$\tau^{\textrm{I}}_{y0}$ (MPa) & Experimental 300 K~\cite{ando2010deformation} & 53\\
 \hline
$\Delta G_{XSI}$ (eV) & MEAM~\cite{WuCurtinScience2018}  & 0.23\\
\hline
$l_{nuc}$ (nm)  & MEAM~\cite{WuCurtinScience2018}  & 2\\
 \hline
$\phi$ & Experimental strong basal texture & $90^\circ$\\
\hline
$\mu$ (GPa) & DFT computed  & 19.5\\
 \hline
$\nu$  & DFT computed  & 0.27\\
 \hline
 $\tau_b$ (MPa) & Related to experimental strain hardening~\cite{Rikihisa2017,Rikihisa2020} &  12 \\
\hline
\hline
 \end{tabular}

\end{center}
    \caption{Values of pure Mg and solute type independent inputs to the cross-slip line-tension  model. Isotropic elastic constants are based on Voigt average~\cite{Hirth1982}.}    
        \label{tab:modelInputsPureMg}
\end{table}

\begin{table}[htbp!]
\footnotesize
\begin{center}
 \begin{tabular}{|c|c|c|}
\hline
Input & Source/Remark &  Value  \\
\hline
\hline
 $\textrm{Avg}_{\textrm{slope}}(\Delta E^{\textrm{I-II}}_{\textrm{Mg-Y}})$ (meV/nm/c) & DFT informed ($\Delta U_j^{\textrm{I-II}}$) &  5100  \\
 \hline

 $\textrm{Fluc}(\Delta E^{\textrm{I-II}}_{\textrm{Mg-Y}})$ (meV/$\sqrt{c}$) & DFT informed ($\Delta U_j^{\textrm{I-II}}$) & 915  \\
 \hline
 $\tau^{\textrm{II}}_{ys-\textrm{Y}}$ (c) at 300 K & DFT informed  ($U_j^{\textrm{II}}$) & Fig.~\ref{fig:panel2} (D) \\ 
 \hline
 $\tau^{\textrm{I}}_{ys-\textrm{Y}}$ (c) at 300 K & DFT informed ($U_j^{\textrm{I}}$) & Fig.~\ref{fig:panel2} (D)  \\
\hline
 $\phi$ & Experimental  & Weakened basal texture: $75^\circ$\\
\hline
 \hline
 \end{tabular}
\end{center}
    \caption{Values of Mg-Y inputs to the cross-slip line-tension model.}    
        \label{tab:modelInputsMgY}
\end{table}

\begin{figure}[htp]
 \centering
 \includegraphics[width=\textwidth]{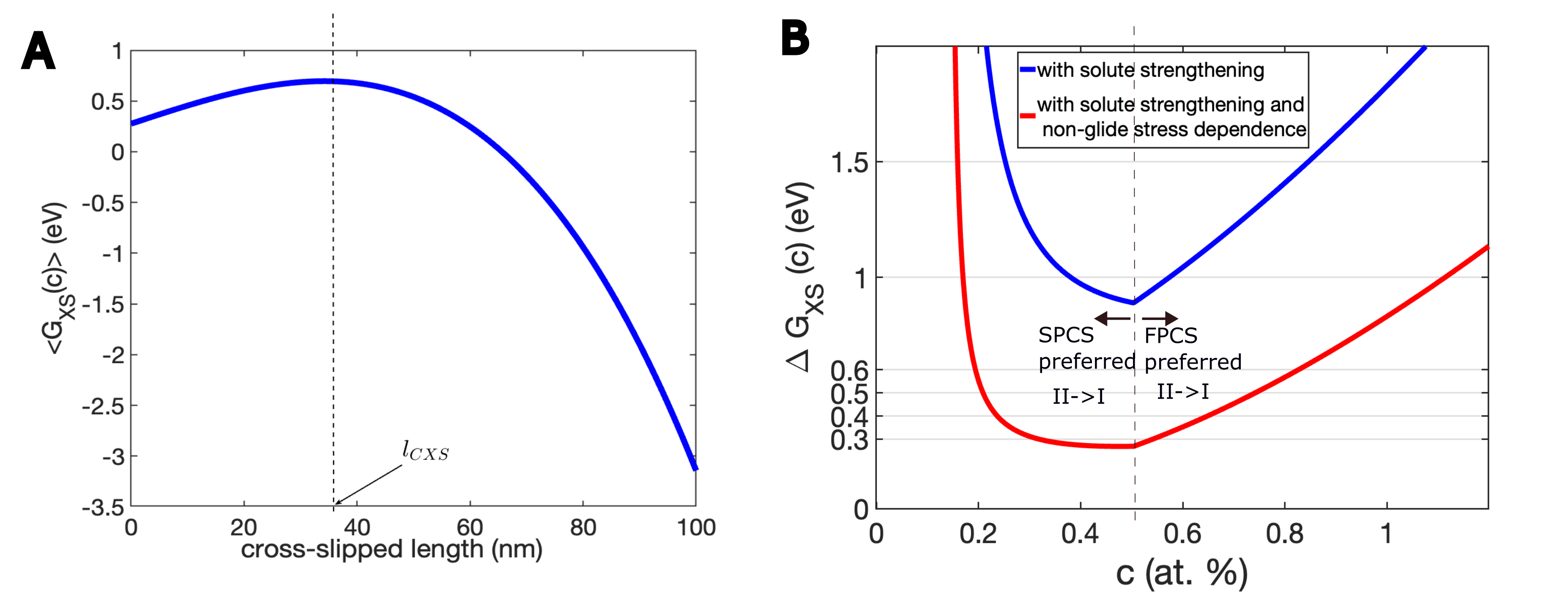}
  \caption{(A) Maximization of average cross-slip energy barrier $\langle \Delta G_{XS}(c)\rangle$  with respect to cross-slipped length in the  dislocation line tension model (cf. Eq.~\ref{eq:crossslipmodelimproved}). Applied to Mg-Y single crystal at 0.4 at.\% Y.  \deltaEpyr~dependency on $\boldsymbol{\epsilon}$ is included; (B) Influence of incorporation of solute-strengthening and macroscopic deformation-core energy dependence  on $\Delta G_{XS}(c)$ prediction (Eq.~\ref{eq:crossslipmodelfloorcondition}) in Mg-Y single crystal. }  
   \label{fig:deltaGBreakdown}
\end{figure}
We now discuss the results of the cross-slip model applied to Mg-Y. We list all the inputs to the cross-slip model in Tabs.~\ref{tab:modelInputsPureMg} and~\ref{tab:modelInputsMgY} for pure Mg and Mg-Y respectively. 
First, we assess the progressive influence of solute strengthening and macroscopic strain dependence of \deltaEpyr~on $\Delta G_{XS}(c)$ in Mg-Y single crystals undergoing tensile loading along $\hkl<1 1 -2 0>$.    $\Delta G_{XS}(c)$ is obtained by maximization of the energy change as function of cross-slipped length in the line-tension model (cf. Fig.~\ref{fig:deltaGBreakdown}(A)). 
Considering the role of solute strengthening effects in Fig.~\ref{fig:deltaGBreakdown}(B), we predict a transition from SPCS to FPCS beyond  0.5 at.\%Y in the tensile loading of Mg-Y single crystal, which is consistent with experimental observations of significant FPCS activity beyond 0.5 at.\%Y~\cite{Rikihisa2017}.
Remarkably, there is a substantial reduction of $\Delta G_{XS}$ to $\sim$0.9 eV at 0.5 at.\%Y compared to smaller concentration values.   This counterintuitive solute strengthening mediated cross-slip barrier reduction can be understood from the concentration dependent CRSS of Pyr I ($\tau_{CRSS}^{I}(c)$) and Pyr II ($\tau_{CRSS}^{II}(c)$). To elaborate, Pyr II CRSS is lower than Pyr I CRSS in pure Mg, but increases at a faster rate compared to that of Pyr I CRSS (cf. Fig.~\ref{fig:panel2}(D)). Using Eq.~\ref{eq:quasistaticEqCondition} in the regime of SPCS preference ($c<0.5$ at.\%Y), the net resolved shear stress for cross-slip on the Pyr I plane (the rate determining step) is 
\begin{equation}
\Delta \tau_{RSS}^{I}(c)= \tau_b \frac{m_{I}(\phi)}{m_{II}(\phi)} +\left(\tau^{II}_{CRSS}(c) \frac{m_{I}(\phi)}{m_{II}(\phi)}-\tau^{I}_{CRSS}(c)\right)\,.\label{eq:DeltaTauRSSI}
\end{equation}
The concentration dependent contribution to $\Delta \tau_{RSS}^{I}(c)$ (the second term in Eq.~\eqref{eq:DeltaTauRSSI}) increases monotonically with $c$, as Pyr II CRSS increases at a faster rate with Y concentration in comparison to Pyr I CRSS.      
Consequentially, a higher value of $\Delta \tau^I_{RSS} (c)$ leads to a lower $\Delta G_{XS}(c)$, and cross-slip enhancement, as evident from Eq.~\ref{eq:crossslipmodelimproved} (cf. ~\ref{sec:appCrossSlipModel}). The next consideration is the dependency of $\Delta E^{\textrm{I-II}}_{\textrm{Mg}}$ on non-glide macroscopic strain/stress. 
Incorporating the macroscopic strain dependence of $\Delta E^{\textrm{I-II}}_{\textrm{Mg}}$ further reduces the $\Delta G_{XS}$ at 0.5~at.\%Y in single crystal Mg-Y to 0.3~eV from 0.9~eV (cf. Fig.~\ref{fig:deltaGBreakdown}(B)), which is below the detrimental  pyramidal to basal transformation barrier of 0.5~eV for edge segments~\cite{WuCurtinScience2018}.  The reduction in $\Delta G_{XS}$ results from a decrease in \deltaEpyr~by accounting for the dependence on strain, especially the tensile strain normal to the Pyr II plane~resulting from the tensile loading along $\hkl<1 1 -2 0>$ direction ~(cf.  Fig.~\ref{fig:deltaEMacroSolute}).

\begin{figure}[htp]
 \centering
 \includegraphics[width=\textwidth]{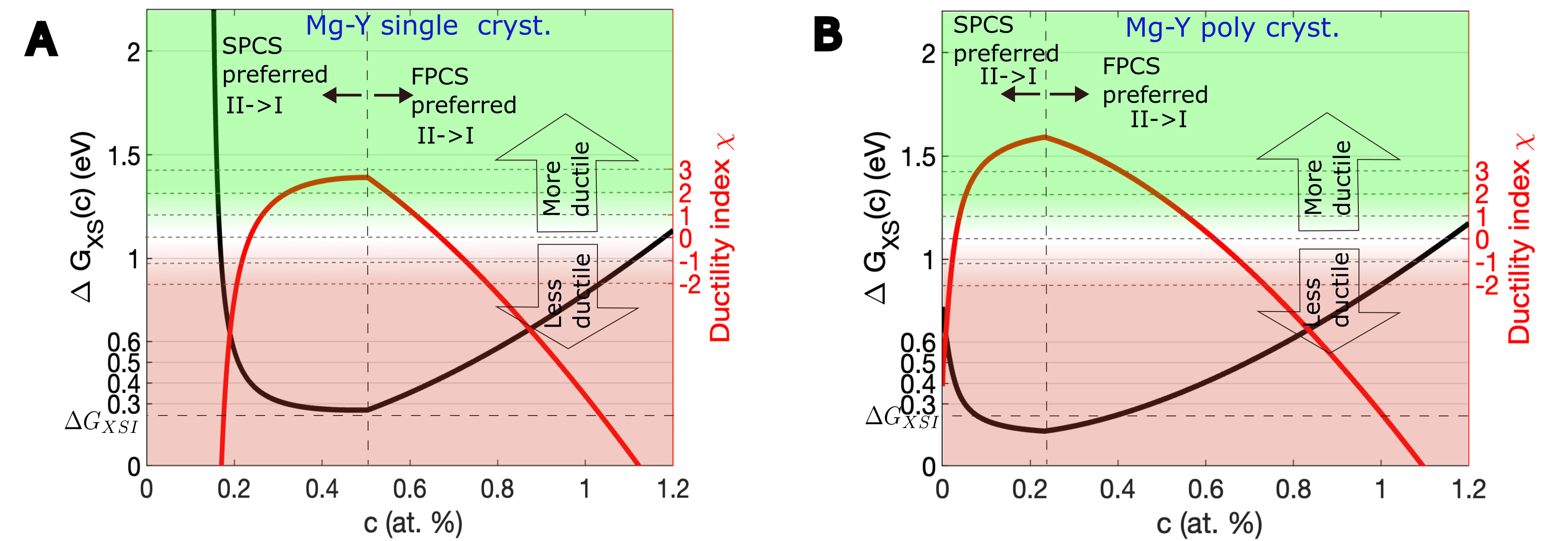}
   \caption{Cross-slip barrier ($\Delta G_{XS}$) and ductility index ($\chi$) predictions in (A) Mg-Y single crystal; and (B) Mg-Y poly crystal. Preferred slip (SPCS/FPCS) and rate-limiting cross-slip transformations (Pyr II$\rightarrow$Pyr I or Pyr I$\rightarrow$Pyr II) are indicated. $\Delta G_{XSI}$ is the barrier for cross-slip nucleation.}  
   \label{fig:panel3}
\end{figure}

The non-monotonic nature of $\Delta G_{XS}(c)$, as depicted in Fig~\ref{fig:panel3}(A), can be understood as follows. The first source of non-monotonicity is the change in the primary slip system from SPCS to FPCS at $\sim$ 0.5 at.\% Y, which creates a non-smooth change in the  $\Delta G_{XS}(c)$. The other source of non-monotonicity, that manifest as smooth variation in $\Delta G_{XS}(c)$ within each slip preference regime, arise from the various competing solute effects that exhibit different scaling with respect to solute concentration. These are: (i) the average solute effect ($\langle\Delta E^{\textrm{I-II}}_{\textrm{Mg-x}}\rangle$) that scales linearly, and increases $\Delta G_{XS}$, with concentration; (ii) the solute effects on $\tau_{CRSS}$ that scales as $c^{2/3}$ (cf. ~\ref{sec:appCRSS}) 
; and (iii) the fluctuation effects that scale as $\sqrt{c}$ and reduces $\Delta G_{XS}$.   
The solute strengthening and fluctuation effects dominate for smaller concentrations.
On the other hand, for larger concentrations, the average solute effect that increases $\Delta G_{XS}$ dominates. This increasing $\Delta G_{XS}$ for larger Y concentrations, which reduces cross-slip activity and hence decreases also the propensity of \cplusa~dislocation multiplication, can also be a reason for experimental observations of reduced \cplusa~ activity beyond 0.9~at\%Y~\cite{Rikihisa2017}. Furthermore, we have applied the cross-slip model to poly-crystalline Mg-Y, where our predictions are shown in Fig.~\ref{fig:panel3}(B). We observe a non-monotonic dependence of $\Delta G_{XS}(c)$ on the concentration similar to the single-crystal Mg-Y case but with a wider zone of ductility enhancement starting from small concentrations of $\sim$0.1 at.\% Y. The SPCS-FPCS transition occurs at $\sim$0.25 at.\%Y, lower than the transition point of $\sim$0.5 at.\%Y in the case of single crystal Mg-Y. This is primarily a consequence of the increased value of the averaged Schmidt factor of Pyr I plane relative to Pyr II plane in the poly-crystalline case. These results are in line with recent experimental results on rolled Mg-Y poly-crystalline sheets~\cite{Rikihisa2020}, where non-monotonic ductility enhancement is observed.

Using the computed cross-slip barriers for both single-crystal and poly-crystal Mg-Y, we follow~\cite{WuCurtinScience2018} to compare the  \cplusa~screw cross-slip reaction rate in relation to the detrimental pyramidal to basal (PB) transformation rate of edge segments at 300~K, and denote the ductility index as $\chi=(\textrm{ln}(l_{PB}/l_{XS})+(\Delta G_{PB}-\Delta G_{XS}(c))/(k_BT))/\left(\textrm{ln}(10)\right)$. $l_{XS}$ and $l_{PB}$ denote the critical nucleation lengths for the cross-slip and PB transition, respectively. We use values from~\cite{WuCurtinScience2018} of $l_{PB}=2\,\textrm{nm}$ and $\Delta G_{PB}=0.5\,\textrm{eV}$. 
We note that when the ductility index is greater than 1, the cross-slip of screw components is more than 10-fold faster than the detrimental PB transition of edge segments. Figures~~\ref{fig:panel3}(A)-(B)  show the ductility index for single- and poly-crystal Mg-Y, where enhanced ductility is observed for concentrations in the range of  0.2--0.7 and  0.1--0.6  at.\% Y respectively.

\subsection{Cross-slip energy barriers and ductility enhancement predictions for Mg-Zn}
\begin{table}[htbp!]
\footnotesize
\begin{center}
 \begin{tabular}{|c|c|c|}
\hline
Input & Source/Remark  & Value\\
\hline
\hline
 $\textrm{Avg}_{\textrm{slope}}(\Delta E^{\textrm{I-II}}_{\textrm{Mg-Zn}})$ (meV/nm/c) & DFT informed ($\Delta U_j^{\textrm{I-II}}$)   & -10050\\
 \hline

 $\textrm{Fluc}(\Delta E^{\textrm{I-II}}_{\textrm{Mg-Zn}})$ (meV/$\sqrt{c}$) & DFT informed ($\Delta U_j^{\textrm{I-II}}$)   & 1431 \\
 \hline
 $\tau^{\textrm{II}}_{ys-\textrm{Zn}}$ (c) at 300 K & DFT informed  ($U_j^{\textrm{II}}$) & Fig.~\ref{fig:panel2} (D)  \\ 
 \hline
 $\tau^{\textrm{I}}_{ys-\textrm{Zn}}$ (c) at 300 K & DFT informed ($U_j^{\textrm{I}}$)  & Fig.~\ref{fig:panel2} (D) \\
\hline
 $\phi$ & Experimental   & Strong basal texture: $90^\circ$ \\
\hline
 \hline
 \end{tabular}
\end{center}
    \caption{Values of Mg-Zn inputs to the cross-slip line-tension model.}    
        \label{tab:modelInputsMgZn}
\end{table}
We also considered Mg-Zn alloys, and estimated the cross-slip barrier as a function of Zn concentration. The tensile loading in the cross-slip model is chosen to be consistent with experiments~\cite{ando2012plasticMgZn,shi2013effects}. In case of single-crystal Mg-Zn, the tensile loading direction is chosen to be along $\hkl<1 1 -2 0>$ direction. In case of poly-crystalline Mg-Zn, the loading direction is chosen to be along the basal plane corresponding to strong basal texture ($\phi=90^{\circ}$), as only mild texture weakening is observed experimentally~\cite{shi2013effects}. The relevant inputs to the cross-slip model for Mg-Zn are listed in Tab.~\ref{tab:modelInputsMgZn}. We note that there are two key qualitative differences in the cross-slip mechanism in Mg-Zn  compared to the Mg-Y,  related to the different behavior of  $\langle\Delta E^{\textrm{I-II}}_{\textrm{Mg-x}}(\boldsymbol{\epsilon})\rangle$ and the solute-strengthening effects on the Pyr \cplusa~screw dislocations. First, there is a sign reversal of $\langle\Delta E^{\textrm{I-II}}_{\textrm{Mg-x}}(\boldsymbol{\epsilon})\rangle$ from positive to negative around $\sim$ 0.05 at. \% Zn   whereas $\langle\Delta E^{\textrm{I-II}}_{\textrm{Mg-x}}(\boldsymbol{\epsilon})\rangle$ is positive at all concentrations for  Mg-Y  (cf. Fig.~\ref{fig:deltaEMacroSolute}). The sign reversal leads to change in the rate determining step from Pyr II$\rightarrow$Pyr I cross-slip at Zn concentrations less than 0.05 at.\% to   Pyr I$\rightarrow$Pyr II cross-slip at  concentrations beyond 0.05 at.\% , effectively the entire dilute concentration range of interest. Second, as discussed in Sec.~\ref{sec:incorporation}, the Pyr I screw CRSS in Mg-Zn is predicted to be higher than the Pyr II screw CRSS unlike the CRSS cross-over in Mg-Y (cf. Fig.~\ref{fig:panel2} (D)). As a consequence, using  the quasi-static loading condition in Eq.~\ref{eq:quasistaticEqCondition}, we find that the SPCS slip is preferred in Mg-Zn with no switch over to FPCS slip preference with increasing Zn concentration. This aligns with experimental observation of SPCS slip during single crystal Mg-Zn plastic deformation~\cite{ando2012plasticMgZn}. The above   aspects pertaining to the Mg-Zn double cross-slip mechanism are schematically illustrated in Fig.~\ref{fig:lineTensionModelSchematicMgZn}.

\begin{figure}[htp]
 \centering
 \includegraphics[width=\textwidth]{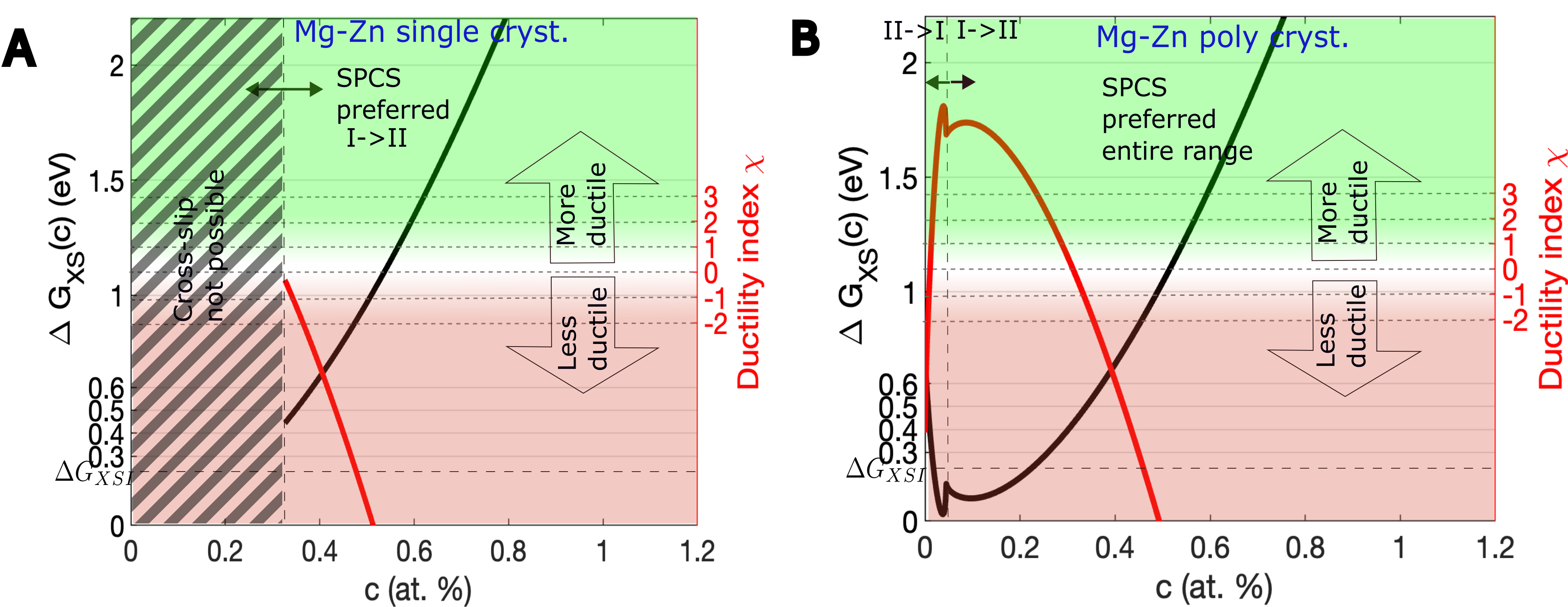}
  \caption{Cross-slip barrier and ductility index predictions in (A) Mg-Zn single crystal; and (B) Mg-Zn poly-crystal. Preferred slip planes and rate-limiting cross-slip transformations are indicated in the various dilute concentration regimes. Concentration regime where cross slip is not possible due to non-satisfaction of the thermally activated glide condition (cf. Eq.~\ref{eq:thermallyActivatedGlide}) is also indicated.}  
   \label{fig:doubleCrossSlipMechanismDuctilityPredictionsMgZn}
\end{figure}

Next, we discuss the $\Delta G_{XS}(c)$ predictions computed using the cross-slip line-tension model  (cf.~\ref{sec:appCrossSlipModel})  for single- and poly-crystal  Mg-Zn,  plotted in Figs.~\ref{fig:doubleCrossSlipMechanismDuctilityPredictionsMgZn} (A) and (B). First, considering the single-crystal, we observe that the $\Delta G_{XS}(c)$ attains small values only in a small concentration window beyond $\sim$ 0.3 at. \% Zn. Under 0.3 at.\% Zn, the resolved shear stress on the non-preferred Pyr I plane is not large enough to enable thermally activated glide, a necessary condition for double cross-slip activation (using Eq.~\ref{eq:thermallyActivatedGlide}).   On the other hand, at larger Zn concentrations, the average solute effect that linearly increases the magnitude of $\langle\Delta E^{\textrm{I-II}}_{\textrm{Mg-x}}\rangle$  results in a monotonically increasing $\Delta G_{XS}(c)$. Moving to the poly-crystalline case, we note that the averaging over the strongly basal poly-crystal texture leads to an increase in the Schmidt factor for the Pyr I plane relative to Schmidt factor of the Pyr II plane, thereby increasing the resolved shear stress on Pyr I plane exerted by the SPCS controlled quasi-static loading. This facilitates the satisfaction of the thermally activated glide condition  and prediction of low  $\Delta G_{XS}(c)$ values (less than 0.3 eV) even at small Zn concentrations  $\sim$ 0.1 at.\%. Overall, we predict mild ductility enhancement in Mg-Zn single crystal and larger enhancement for poly-crystal Mg-Zn between 0.05--0.5 at.\% Zn, in line with room-temperature tensile test experimentation results~\cite{shi2013effects,ando2012plasticMgZn,JavierLLorca2024}.

\section{Conclusions}
The main conclusions from our study are:
\begin{enumerate}
    \item We have computed the cell size converged dislocation and dislocation-solute energetics of \cplusa~screw dislocations in pure Mg and dilute Mg alloys from large-scale DFT calculations. We demonstrate substantial qualitative differences in predictions of solute effects on the \cplusa~core energy difference obtained from explicit DFT calculations in this work compared to prior estimates using indirect stacking fault-solute interaction energies.
    \item Subsequently, we proposed a first-principles informed  model for the critical energy barrier prediction in the \cplusa~double cross-slip mechanism, that incorporates solute-strengthening effects and effect of macroscopic deformation on core energetics. We find the solute-strengthening effects to be the key driver for cross-slip enhancement in Mg-Y. Our modeling successfully explains experimental observation of enhanced room temperature ductility for Mg-Y and the associated SPCS to FPCS transition. 
    \item Our model also explains the suppressed room temperature ductility in single-crystal Mg-Zn resulting from the different nature of solute-strengthening effects compared to Mg-Y.
\end{enumerate}
The  cross-slip  mechanism proposed in this work paves the way for improving ductility in general binary and ternary dilute Mg alloys by leveraging the two dimensional control space of the average solute effects and solute strengthening effects using direct first-principles calculations. In this context, ternary Mg alloys~\cite{masood2022designing}, such as Mg-Ca-Zn~\cite{HA2021111044,chino2010enhancement} have emerged as strong alternatives for Mg-RE alloys, and present fruitful directions for future research. We anticipate that the recent advent of exascale computing machines in conjunction with large-scale DFT frameworks, such as the one employed in this work~\cite{MOTAMARRI2013308,das2022dft}, will be critical in accelerating the search for highly ductile Mg alloys. Finally, while the proposed intrinsic ductility model lacks physics of solute-effects on grain-boundaries and microstructure evolution, the model can be integrated into higher-scale crystal-plasticity frameworks developed for Mg alloys~\cite{CPFEMgY,YAGHOOBI2019109078}, that capture the polycrystalline deformation behavior.

\appendix

\section{CRSS prediction of \cplusa~pyramidal screw dislocations using Labusch type solute strengthening}\label{sec:appCRSS}
We briefly describe the Labusch type weak pinning approach~\cite{leyson2010quantitative,leyson2012solute} to predict the solute strengthening of \cplusa~Pyr screw dislocation glide in dilute Mg alloys. A straight dislocation line in a random environment of solutes can reduce its energy by assuming a roughened or wavy configuration to avail favourable solute fluctuations. The wavy configuration has a characteristic roughening length $\zeta$ and roughening amplitude $w$ (cf. schematic in Fig.~\ref{fig:lineTensionModelSchematicMgY}). Under the assumption that $w$ is much smaller than $\zeta$, the energy change per unit length of the roughened dislocation segment  with respect to the straight dislocation is~\cite{leyson2012solute}
\begin{equation}\label{eq:totalenergyRoughenedSegment}
    \Delta E_{tot}(\zeta,w)=\left[\frac{\Gamma w^2}{2\zeta}-{\left(\frac{c\zeta}{a_p}\right)}^{1/2}\Delta \tilde{E}_p(w)\right] \frac{L}{2\zeta}\,,
\end{equation}
where the first term quantifies the increase in the elastic energy due to the wavy texture, the second term quantifies the reduction in the potential energy due to favorable energy fluctuations from dislocation-solute interactions, and $a_p$ is the periodic length along the dislocation line direction. In the above, $\Gamma$ is the dislocation line tension,   approximated as $\mu \frac{b^2}{4}$ for the straight screw dislocation. At dilute concentrations, $\Delta \tilde{E}_p(w)$ is statistically determined to be dependent on the dislocation-solute interaction energy map  $U(x_i,y_j)$ as
\begin{equation}
    \Delta \tilde{E}_p(w)={\left[\sum_{ij}{\left(U(x_i,y_j)-U(x_i-w,y_j)\right)}^2\right]}^{1/2}\,,
\end{equation}
where $(x_i,y_j)$ is the position of the solute site with respect to the dislocation at the origin. The coordinate $x_i$ is in the slip direction along the glide plane and $y_j$ is normal to the glide plane.  The term ${\left(\frac{c\zeta}{a_p}\right)}^{1/2}\Delta \tilde{E}_p(w)$ term in Eq.~\ref{eq:totalenergyRoughenedSegment} is the standard deviation  of the energy change associated with sampling $U$ when a dislocation segment of length $\zeta$ glides over a distance $w$ along the slip plane. We note that unique solute sites per unit periodic length need not lie on a common $x-y$ plane. Since $\zeta$ is assumed to be much larger than $w$ and $a_p$, we project the 3D site coordinates on to the $x-y$ plane to obtain the $U(x_i,y_j)$ map. 
Next we minimize $\Delta E_{tot}(\zeta,w)$ with respect to $\zeta$ (analytically) and $w$ (numerically) to obtain the critical roughening length $\zeta_c$ and critical roughening amplitude $w_c$.  The dislocation roughening related energy barrier for each segment of length $\zeta_c$ is obtained as~\cite{leyson2012solute}
\begin{equation}\label{eq:crssLabusch1}
\Delta E_b=\beta_0 {\left(\frac{c \sqrt{3} w_c^2 \Gamma \Delta \tilde{E}_p^2(w_c)}{a_p}\right)}^{1/3}\,,
\end{equation}
where $\beta_0$ is an unitless geometric factor dependent on the dislocation slip system. The value of $\beta_0$ for the Pyr \cplusa~ screw dislocation systems computed in the present study is 1.4668. Assuming a sinusoidal variation of the energy landscape, the zero temperature additional resolved shear stress required to be applied for barrier less dislocation glide is obtained as~\cite{leyson2012solute}
\begin{equation}\label{eq:crssLabusch2}
\tau_{ys-\textrm{x}}(c,T=0)=\frac{\pi}{2}\frac{\Delta E_b}{b \zeta_c w_c}\,,
\end{equation}
where $b$ is the dislocation Burgers vector magnitude. Further considering a constant strain rate dislocation flow at finite temperature a lower energy barrier is determined from transition state theory and Kock's law~\cite{leyson2012solute}
\begin{equation}\label{eq:crssLabuschStrainRateTemperature}
\tau_{ys-\textrm{x}}(c,T,\dot{\epsilon})= \tau_{ys-\textrm{x}}(c,T=0)\left(1-{\left(\frac{kT}{\Delta E_b}\textrm{ln}\frac{\dot{\epsilon}}{\dot{\epsilon_0}}\right)}^{2/3}\right)
\end{equation}
where $\dot{\epsilon_0}$ is the reference strain rate, and $\dot{\epsilon}$ is the macroscopic strain rate due to external loading.  We note that $\tau_{ys-\textrm{x}}(c,T=0)$ scales  as $c^{2/3}$ after factoring in the concentration dependence of both  $\Delta E_b$ and  $\zeta_c(w)={\left(\frac{4 \Gamma^2 w^4 a_p}{c\Delta\tilde{E}_p^2(w)}\right)}^{1/3}$,  the latter obtained from the analytical minimization of $\Delta E_{tot}(\zeta,w)$ (cf. Eq.~\ref{eq:totalenergyRoughenedSegment}) with respect to $\zeta_c$~\cite{leyson2012solute}. Further, we consider $\dot{\epsilon_0}=10^{5} \,{\textrm{s}}^{-1}$ as typically considered in that range~\cite{leyson2010quantitative,fellinger2022solutes}, and $\dot{\epsilon}=5\times 10^{-5}\,{\textrm{s}}^{-1}$ corresponding to the relevant tensile loading experiments on Mg alloys~\cite{Rikihisa2017,Rikihisa2020}. Fig.~\ref{fig:panel2} (D) plots the Y and Zn solute concentration dependent CRSS (pure Mg CRSS plus $\tau_{ys-\textrm{x}}$) for both the Pyr I and Pyr II screw dislocation glide. The baseline CRSS values for pure Mg ($c=0$)  Pyr II slip (40 MPa) and Pyr I slip (53 MPa) are obtained from single crystal tensile experimental data at $\sim$ 300 K~\cite{ando2010deformation}. 

We remark that the primary contribution to Pyr slip CRSS in Mg at room temperature is considered to arise from the screw component (35--45 MPa) compared to the edge components (6--8 MPa), calculated using MEAM MD simulations at 300 K~\cite{Wu_2015}. We expect the gap between the edge and screw Pyr CRSS to hold under solute strengthening in dilute alloys, and thus focus only on the solute strengthening prediction for Pyr screw glide in dilute Mg alloys.

\section{Incorporation of \deltaEpyr~dependency on external macroscopic stress}\label{sec:appMacro}
In Sec.~\ref{sec:resultsMacroDeltaEpyr}, we computed the influence of  external uniaxial macroscopic strains on \deltaEpyr~from DFT calculations. In order to evaluate \deltaEpyrmacro~for an arbitrary external macroscopic strain state $\boldsymbol{\epsilon}$ as required in the line-tension model (cf. Eq.~\ref{eq:crossslipmodelimproved}), we approximate the strain dependency through a linear (piecewise linear) approximation of the core-energy dependence about each strain component. In particular, the piecewise linear approximation is used for the \deltaEpyrmacro's asymmetric dependence on the $\boldsymbol{\epsilon_{22}}$ strain (normal to Pyr II plane) (cf. Tab.~\ref{tab:stressDependency}). The above linear approximation is valid for small strain magnitudes during the onset of yielding in Mg ($\sim$0.2\% strain), the zone of interest for our cross-slip barrier predictions. We note the values of the \deltaEpyrmacro~dependency are given with respect to $\boldsymbol{\epsilon}$ strain tensor in the coordinate frame of the Pyr II slip-plane, whose evaluation we describe below.  Recall from Eq.~\ref{eq:quasistaticEqCondition} in Sec.~\ref{sec:incorporation} that we have access to local quasi-static tensile loading stress $\boldsymbol{\sigma}_y(c)$ along the loading direction, which in turn is dependent on the Pyr plane Schmidt factors,
CRSS values, and texture in case of a poly-crystalline microstructure. Next, the desired strain tensor is evaluated from $\boldsymbol{\sigma}_y(c)$ using Voigt averaged isotropic elastic constants~\cite{Hirth1982}, followed by coordinate transformation to the Pyr II plane aligned coordinate frame in which the  the \deltaEpyrmacro~dependency data is available.

\section{Cross-slip line tension model}\label{sec:appCrossSlipModel}
We  discuss below the expressions for the dislocation line tension model to compute the cross-slip energy barrier. The model is  adapted  from Refs. ~\cite{WuCurtinScience2018,AhmadCurtin2019}. The key improvements in this work are the use of first-principles dislocation energetics inputs as opposed to atomistic inputs, incorporation of solute-strengthening effects in estimation of net resolved shear stress for cross-slip ($\Delta \tau_{RSS}(c)$), and incorporation of dislocation core energy-macroscopic deformation dependence (\deltaEpyrmacro). Recalling the discussion in Sec.~\ref{sec:incorporation}, the rate-determining cross-slip step in the double cross-slip mechanism is the cross-slip event from the lower Pyr screw dislocation energy  plane to the higher Pyr screw dislocation energy  plane. In the case of Mg-Y, the low and high energy planes are Pyr II and Pyr I, respectively, at all dilute concentrations, while in Mg-Zn the stability ordering gets reversed beyond $\sim$ 0.05 at.\% Zn. In the following discussion, we will consider the former case of Mg-Y with the rate-determining step accordingly from Pyr II to Pyr I plane. The same expressions apply to the case when Pyr I$\rightarrow$Pyr II cross-slip is the rate determining step, such as in Mg-Zn, after considering the change in the sign of the energetics inputs. 
In the line tension model, the concentration dependent  cross-slip energy barrier, denoted by $\langle\Delta G_{XS}(c)\rangle$, is obtained  by maximizing the following energy change with respect to the cross-slipped length $l$:
\begin{align}\label{eq:crossslipmodelimproved}
 \langle\Delta G_{XS}(c)\rangle(l) = & \Delta G_{XSI}  + \langle\Delta E^{\textrm{I-II}}_{\textrm{Mg-x}}(\boldsymbol{\epsilon})\rangle \left(l_{nuc}+l\right)  \notag\\ & + \Gamma \Delta s(l) - \Delta \tau_{RSS} (c) b A(l),\,\quad \left.\frac{\partial \langle\Delta G_{XS}(c)\rangle(l)}{\partial l}\right|_{l=l_{CXS}} = 0\,,
\end{align}
where ``$\langle \rangle$''  denotes the consideration  of average solute effects as will be discussed below. In the above,  
 $\Delta G_{XSI}$ is the intrinsic barrier associated with the initial cross-slip nucleation on the Pyr I plane, $\langle\Delta E^{\textrm{I-II}}_{\textrm{Mg-x}}(\boldsymbol{\epsilon})\rangle$ denotes the average dislocation core energy difference per unit length between Pyr I  and Pyr II screw dislocations accounting for solute-effects and influence of external macroscopic strain,  $\Gamma$ is the \cplusa~screw dislocation line tension, $\Delta s(l)$ is the increase in the dislocation line length due to bowing out of the cross-slipped dislocation line on the higher energy Pyr plane,  $\Delta \tau_{RSS}(c)$ is the concentration dependent net resolved shear stress available for the cross-slip process after subtracting the critical resolved shear stress $\tau_{CRSS}(c)$ on the cross-slip plane, $b$ is the Burgers vector magnitude of the \cplusa~screw dislocation, and $A(l)$ is the area swept on the pyramidal I plane by the bowed out dislocation line. $l_{CXS}$ is the critical cross-slip length which maximizes $\langle\Delta G_{XS}(c)\rangle(l)$. The radius of the bowed out dislocation segment is obtained from the equilibrium condition involving the line-tension and the Peach-Koehler force exerted by $\Delta \tau_{RSS}$.   
 
Next, we briefly describe the quantification of the solute effects. $\langle\Delta E^{\textrm{I-II}}_{\textrm{Mg-x}}(\boldsymbol{\epsilon})\rangle$ denotes average value of $\Delta E^{\textrm{I-II}}_{\textrm{Mg-x}}(\boldsymbol{\epsilon})$ in a dilute random solute environment, quantified from dislocation-solute interactions as will be discussed below.
Subsequently, we consider the  effect of fluctuation effect created by dislocation-solute interactions, denoted by \sigmaFluc, which quantifies the standard deviation of the $\Delta E^{\textrm{I-II}}_{\textrm{Mg-x}}(\boldsymbol{\epsilon})$ fluctuations about the average effect due to the actual random occupancy of the solutes along a long dislocation line segment of length $l_{CXS}$. The fluctuation effect lowers the energy barrier from the average energy barrier $\langle\Delta G_{XS}(c)\rangle$ at the critical cross-slip length $l_{CXS}$ by one standard deviation. Accordingly, we subtract \sigmaFluc from $\langle\Delta G_{XS}(c)\rangle$ and consider our final $\Delta G_{XS}(c)$ prediction to be the maximum between the resulting value and the intrinsic cross-slip energy barrier $\Delta G_{XSI}$:
\begin{align}\label{eq:crossslipmodelfloorcondition}
  \Delta G_{XS}(c)= & \textrm{max} \{\langle\Delta G_{XS}(c)\rangle- \sigma\left(\mathcal{F}\left[\Delta E^{\textrm{I-II}}_{\textrm{Mg-x}}(l_{CXS})\right]\right) \,;\,\,\Delta G_{XSI}\}\,.
\end{align}
In the above, the average and fluctuation effects are quantified using statistical analysis of the site dependent \deltaUpyr, defined as the difference of the dislocation-solute interaction energies between the Pyr I and II dislocations  at an  substitutional site $j$ ($U_j^I-U_j^{II}$). We note that  higher order solute-solute interactions are neglected in the statistical analysis,  considering dilute concentrations  as detailed in  Refs.~\cite{WuCurtinScience2018,AhmadCurtin2019}. Also the analysis determines that while $\langle\Delta E^{\textrm{I-II}}_{\textrm{Mg-x}}\rangle$ scales linearly with respect to solute concentration $c$ with slope denoted by $\textrm{Avg}_{\textrm{slope}}(\Delta E^{\textrm{I-II}}_{\textrm{Mg-x}})$,  \sigmaFluc~scales as $c^{1/2}$. The relevant expressions are
\begin{align}
    \langle\Delta E^{\textrm{I-II}}_{\textrm{Mg-x}}\rangle
     =&\Delta E^{\textrm{I-II}}_{\textrm{Mg}} + c  \textrm{Avg}_{\textrm{slope}}(\Delta E^{\textrm{I-II}}_{\textrm{Mg-x}}) \notag\\
    \sigma\left(\mathcal{F}\left[\Delta E^{\textrm{I-II}}_{\textrm{Mg-x}}(l_{CXS})\right]\right)= &c^{1/2}\sqrt{\frac{l_{CXS}}{b} } \textrm{Fluc}(\Delta E^{\textrm{I-II}}_{\textrm{Mg-x}})\,,
\end{align}
where the concentration independent $\textrm{Avg}_{\textrm{slope}}(\Delta E^{\textrm{I-II}}_{\textrm{Mg-x}})$ and $\textrm{Fluc}(\Delta E^{\textrm{I-II}}_{\textrm{Mg-x}})$ quantities are computed from  \deltaUpyr~as follows
\begin{align}
    \textrm{Avg}_{\textrm{slope}}(\Delta E^{\textrm{I-II}}_{\textrm{Mg-x}})=& \frac{1}{b}\left( \sum_{j \in N_s} \Delta U_j^{\textrm{I-II}} \right) \notag\\
     \textrm{Fluc}(\Delta E^{\textrm{I-II}}_{\textrm{Mg-x}})=& \left[ \sum_{j \in N_s} {\left(\Delta U_j^{\textrm{I-II}}\right)}^2 \right]\,.
\end{align}
In the above, $j$ sums over a finite number of $N_s$ near core sites per unit $b$ along the dislocation line.  The first-principles  inputs to the line tension model are  $ \tau_{CRSS} (c)$,  \deltaEpyrmacro, and \Us, which are either directly obtained or informed from large-scale DFT calculations in this work. We refer to separate sections in the Appendix for details on modeling  $\tau_{CRSS} (c)$ via Labusch type solute-strengthening model using \Us~inputs, and $\Delta E^{\textrm{I-II}}_{\textrm{Mg}}(\boldsymbol{\epsilon})$ via a linearization simplification. The values of the remaining parameters in the line tension model: $\Delta G_{XSI}, l_{nuc}$ are taken from Refs.~\cite{WuCurtinScience2018,AhmadCurtin2019}, wherin MEAM calculations were performed to obtain these quantities. Further, we approximate the line-tension $\Gamma$ as $\mu \frac{b^2}{4}$, where $\mu$ denotes the shear modulus computed from DFT (cf. Tabs.~\ref{tab:elasticLatticeConstantsDFT} and ~\ref{tab:modelInputsPureMg}).





\bibliography{scibib}

\bibliographystyle{elsarticle-num}

\section*{Acknowledgments}
We gratefully acknowledge the support from the Department of Energy, Office of Basic Energy Sciences (Award number DE-SC0008637 as part of the Center for Predictive Integrated Structural Materials Science) that funded this work. This research used resources of the Oak Ridge Leadership Computing Facility, which is a DOE Office of Science User Facility supported under Contract DE-AC05-00OR22725. This research used resources of the National Energy Research Scientific Computing Center, a DOE Office of Science User Facility supported by the Office of Science of the U.S. Department of Energy under Contract No. DE-AC02-05CH11231. 
\paragraph{Author contributions:} Conceptualization: SD, VG;
Methodology: SD, VG;
Investigation: SD, VG;
Visualization: SD;
Funding acquisition: VG;
Supervision: VG;
Writing-original draft: SD, VG;
Writing-review \& editing: SD, VG.

\paragraph{Competing interests:} Authors declare that they have no competing interests.

\paragraph{Data and materials availability:} We have deposited the raw input and output files for the ab-initio calculations, performed using the DFT-FE code, into NOMAD public repository. The deposited files can be found at this \href{https://nomad-lab.eu/prod/v1/gui/search/entries/entry/id/s9wj7W5LuwPsu3x05P5NyjephXAm}{link}.


\clearpage

\end{document}